\documentclass{aa}

\usepackage{graphicx}
\usepackage{txfonts}
\usepackage{natbib}
\usepackage[figuresright]{rotating}
\usepackage{subfigure}
\usepackage{lscape}
\usepackage{aalongtable}
\usepackage{psfig}
\usepackage{color}
\usepackage{afterpage}
\usepackage{natbib}

\newcommand{\msun}{$M_\odot$}

\def\h2{H{\small II}}

\newcounter{qub}
\begin{document}

   \title{The interplay between ionized gas and massive stars in the HII galaxy IIZw70: integral field spectroscopy with PMAS}
   \titlerunning{HII galaxy IIZw70: integral field spectroscopy with PMAS}

   \author{C. Kehrig
          \inst{1,2}
          \and
          J.M. V\'{\i}lchez\inst{1}
          \and
          S.F. S\'anchez\inst{3}
          \and
          E. Telles\inst{2}
          \and
          E. P\'erez-Montero\inst{4}
          \and
          D. Mart\'{\i}n-Gord\'on\inst{1}
          }

   \offprints{C. Kehrig}

   \institute{Instituto de Astrof\'{\i}sica de Andaluc\'{\i}a (CSIC),
              Apartado 3004, 18080 Granada, Spain\\
              \email{kehrig@iaa.es,jvm@iaa.es}
         \and
             Observat\'{o}rio Nacional,
             Jos\'{e} Cristino, 77, 20.921-400, Rio de Janeiro - RJ, Brazil\\
             \email{kehrig@on.br,etelles@on.br}
         \and
              Centro Astron\'omico Hispano Alem\'an, Calar Alto, CSIC-MPG,
              C/ Jes\'us Durb\'an Rem\'on 2, E-04004 Almeria, Spain\\
             \email{sanchez@caha.es}
         \and
             Laboratoire d'Astrophysique de Toulouse et Tarbes (LATT - UMR 5572), Observatoire Midi-Pyrnes,
             14 avenue E. Belin, 31400 Toulouse, France \email{enrique.perez@ast.obs-mip.fr}
              }

   \date{Received <date>; accepted <date>; Last update \today}

  \abstract
   {}
   {We performed an integral field spectroscopic study for the
   HII galaxy IIZw70 to investigate the interplay between its ionized interstellar medium (ISM) and the massive star formation.}
   {Observations were taken in the optical spectral range from
   $\lambda$3700~\AA~-~6800~\AA~with the Potsdam Multi-Aperture
   Spectrophotometer (PMAS) attached to the 3.5 m telescope at Calar
   Alto Observatory. We created and analyzed maps of spatially
   distributed emission-lines (at different stages of excitation),
   continuum emission, and properties of the ionized ISM (ionization
   structure indicators, physical-chemical conditions, dust
   extinction, kinematics). We investigated the relation of these
   properties to the spatial distribution and evolutionary stage of
   the massive stars.}
{For the first time we have detected Wolf-Rayet (WR) stars in this
galaxy. The peak of the ionized gas emission coincides with both the
location of the maximum of the stellar continuum emission and the WR
bump. The region of the galaxy with lower dust extinction corresponds
to the region that shows the lowest values of velocity dispersion and
radial velocity. The overall picture suggests that the
ISM of this region is being disrupted via
photoionization and stellar winds, leading to a spatial decoupling
between gas+stars and dust clouds. The bulk of dust appears to be located at
the boundaries of the region occupied by the probable ionizing
cluster. We also found that this region is associated to the nebular
emission in HeII$\lambda$4686 and to the intensity maximum of most emission
lines. This indicates that the hard ionizing radiation
responsible for the HeII$\lambda$4686 nebular emission can be related to
the youngest stars. Within $\sim$ 0.4 $\times$ 0.3 kpc$^{2}$ in the central
burst, we derived oxygen abundances using direct determinations of
$T_{\rm e}$[OIII]. We found abundances in the range 12+log(O/H)= 7.65 - 8.05, yielding an error-weighted mean of 12+log(O/H)=7.86 $\pm$
0.05 that has been taken as the representative oxygen abundance for
IIZw70.}
   {}

   \maketitle

   \section{Introduction}
   \label{intro}

The HII galaxies are gas-rich, dwarf systems that have experienced
intense recent or ongoing violent star formation (SF). These objects were identified
for the first time by Haro (1956) and Zwicky (1964), and are
characterized by the presence of giant HII regions that dominate their
observable properties at optical wavelengths. In particular, their
integrated spectra are very similar to those of giant extragalactic
HII regions in spiral galaxies (Sargent \& Searle 1970).  The analysis
of their emission line spectrum indicates that HII galaxies are metal
poor objects (1/40 Z$_{\odot}$ $\lesssim$ Z $\lesssim$ 1/3
Z$_{\odot}$; e.g. Kunth \& Sargent 1983).

   HII galaxies are ideal laboratories for probing the interplay between
   massive SF and the ISM in low metallicity environments.  The
   massive SF process in dwarf galaxies has effects on the
   properties of the surrounding ISM. A large number of massive stars
   (between 10$^{4}$ - 10$^{6}$ solar masses of gas are transformed
   into stars in HII galaxies) are formed almost simultaneously within
   relatively small volumes (e.g. Melnick 1992).
   Millions of years after the onset of the burst, the most massive
   stars begin to explode as supernovae, creating violent,
   short-lived injections of kinetic energy and metallic elements into
   the ISM. The disruption of the ISM may significantly affect the
   spatial distribution of gas and dust particles in the regions close
   to the massive star cluster, and also determine the way in which
   new metals ejected by massive stars are mixed with the original gas
   from which the stars formed (Kunth \& \"Ostlin 2000).

   High spatial resolution imaging has revealed that in many HII
   galaxies the ionized material presents a complex structure: star
   clusters in the main body (with a non-uniform distribution of
   SF knots, ensembles of star clusters, or individual super
   stellar clusters) of the galaxy and external gas components outside
   the young stellar clusters (e.g. Martin \& Kennicutt 1995,
   V\'{\i}lchez \& Iglesias-P\'aramo 1998, Papaderos et
   al. 2002). Thus, a two-dimensional analysis of the ionized material
   in HII galaxies yields spatially resolved information on properties
   of the ionized gas.  The study of the distribution of these
   properties is an important issue for our understanding of the
   interplay between the massive stellar population and the ISM.

   However, up to now only a few two-dimensional spectroscopic studies
   of HII galaxies have been performed. Recent work by Izotov et
   al. (2006) presents two-dimensional spectroscopy of the extremely
   metal-deficient HII galaxy SBS
   0335-052E. They found a small gradient of the electron temperature
   $T_e$ and oxygen abundance, and the presence of an ionized gas
   outflow in the perpendicular direction to the galaxy disk. Cair\'os
   et al. (2002), using two-dimensional spectroscopy, obtained the
   ionized gas velocity field in the central part of Mrk 370.

In this paper, we present a two-dimensional spectroscopic study for
the HII galaxy IIZw70. This galaxy is number 70 in Zwicky's second
list of ``Compact Galaxies and Compact Parts of Galaxies, Eruptive and
Post-eruptive Galaxies''(Zwicky \& Zwicky 1971).  IIZw70 has been
classified as an HII galaxy (Sargent 1970). The basic data of IIZw70
are shown in table~\ref{tab1}. This object is interacting with its
nearby companion IIZw71 at a projected distance of 23 kpc (assuming a
distance to the IIZw70-71 system of 18.1 Mpc; see table 1). The
ongoing interaction is indicated by interferometric \ion{H}{I} studies
(Balkowski et al. 1978, Cox et al. 2001) which revealed a gaseous
streamer connecting IIZw70 with the polar ring galaxy candidate IIZw71
(Whitmore et al. 1990, Reshetnikov
\& Combes 1994). Cair\'os et al. (2001) show that IIZw70 presents very
elongated outer isophotes and a very blue (U-B = -0.89) nuclear
starburst in what seems to be an edge-on disk. They also find that the
starburst activity is concentrated in the optical center of the galaxy
and is surrounded by faint gaseous emission with quite a distorted
morphology.

\begin{figure}
\center{
\includegraphics[width=9cm,clip]{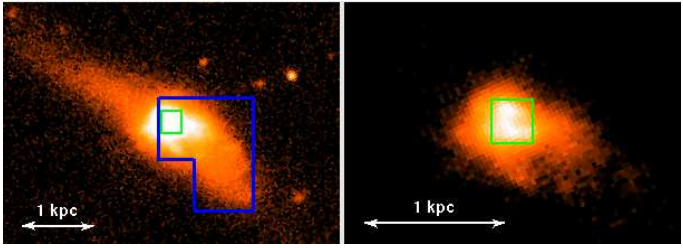}}
\caption{Image of IIZW70 in the {\it r (left)} and {\it u (right)} bands from Sloan Digital Sky
Survey. The blue box  mosaic represents the field used in this work ({\it left}), which is located approximately in the central parts of the galaxy; the green box marks the likely ionizing clusters location, as indicated with a box on the {\it u}  image ({\it right}). North is up, east is left. The spatial scale is shown in each panel, where 1 kpc $\sim$ 11\hbox{$^{\prime\prime }$} assuming a distance to the galaxy of 18.1 Mpc (see table 1).}
\label{SDSS} \end{figure} 

In figure~\ref{SDSS} we show the images of IIZw70 in the {\it r} (left
image) and {\it u} (right image) bands extracted from
SDSS\footnote{The Sloan Digital Sky Survey website is
http://www.sdss.org}. Over the {\it r} filter image, we marked the
field of view used in this work (blue box mosaic).

In this work, we investigate the relation between massive SF process
and the ionized ISM in IIZw70 using integral field spectroscopy
(IFS). In order to do this we created and analyzed maps of spatially
distributed emission-lines, continuum emission and physical-chemical
properties of the ionized gas (electron temperature and density,
gaseous metal abundances, dust extinction, excitation). We present a
discussion on the significance of the chemical abundance variation
found here.  We also present a brief study of the kinematic properties
of the ionized gas.  The relationship between the physical-chemical
and kinematic properties spatial distribution is discussed.

\begin{table}
\caption{Basic data of IIZw70  \label{tab1}}
\renewcommand{\footnoterule}{}
\begin{minipage}{\textwidth}
\begin{tabular}{lc} \hline
Parameter      &Value \\ \hline
Name           &IIZw70  \\
Other designations                &UGC 9560, Mrk 829    \\
R.A. (J2000.0)                 &14h 50m 56.5s         \\
DEC. (J2000.0)                &+35d 34' 18''         \\
redshift   &0.004         \\
M$_{B}$\footnote{Absolute magnitude in B from Cair\'os (2001)}(mag)    &-16.3         \\
m$_{B}$\footnote{Apparent magnitude in B from Deeg et al. (1997)}(mag)  &15.1         \\
D\footnote{Distance to the galaxy from Cox et al. (2001)}(Mpc)       &18.1         \\
M$_{HI}$\footnote{Neutral hydrogen mass from Thuan \& Martin (1981)}(\msun) &0.34 x 10$^{9}$          \\
Z\footnote{Metallicity from Kobulnicky \& Skillman (1996)}/Z$_{\odot}$                          & 1/5  \\
\hline
\end{tabular}
\end{minipage}
\end{table}

In the following section we describe the observations and data
reduction. In Sect.3 the results are presented. We discuss our results
in Sect.4.

\section{Data}

\subsection{Observations}
\label{spectra}

IFS allows us to simultaneously collect spectra of many different
regions of an extended object, under identical instrumental and
atmospheric conditions. This is not only a more efficient way of
observing, but it also guarantees the homogeneity of the dataset.

The spectra of IIZw70 were taken with PMAS that is an integral field
unit (IFU) developed at the Astrophysikalisches Institut Potsdam (Roth et al. 2005). It
is currently installed at the Calar Alto Observatory 3.5 m Telescope.
Two pointings of the galaxy (IIZw70-NE and IIZw70-SW) were taken in order
to cover the central regions and different regions of interest of the galaxy
(e.g. the tails).

The log of observations is given in table~\ref{tab2}.  The V300 and V1200
gratings present a dispersion of 1.6~\AA/pixel and 0.36 ~\AA~/pixel,
respectively. The PMAS
spectrograph is equipped with 256 fibers coupled to a $16\times16$
lens array. Each fiber has a spatial sampling of
$1\hbox{$^{\prime\prime}$}\times1\hbox{$^{\prime\prime}$}$ on the
sky resulting in a field of view of $16\hbox{$^{\prime\prime}$}\times16\hbox{$^{\prime\prime}$}$. 
 Calibration images were obtained following the science
exposures and consisted of emission line lamps spectra (HgNe), and
spectra of a continuum lamp needed to locate the 256 individual
spectra on the CCD. Observations of the spectrophotometric standard
stars BD +28$^\circ$4211 and Hz44 were obtained during the observing
nights for flux calibration. The nights were photometric and the typical
seeing during the observations was 1$^{\prime\prime}$.


\begin{table}
\caption{Log of the observations.}
\label{tab2}
\centering
\begin{tabular}{lcccc}

\hline\hline

Pointing 	& Exptime 	& Grating 	& Spec. range  & Date \\
IIZw70  	&  (s)          &	  	&  (~\AA~) & \\
 &	 & & & \\ \hline
IIZw70-NE   	&3x750 	        &V300 	&3900-7049 &05/08/2005 \\
IIZw70-SW   	&3x750 	        &V300 	&3900-7049 &05/08/2005 \\
IIZw70-NE    	&3x1200 	&V1200 	&4330-5020 &06/08/2005 \\
IIZw70-SW    	&3x1200 	&V1200 	&4330-5020 &06/08/2005 \\
IIZw70-NE 	&3x1200 	&V1200 	&3610-4379 &04/06/2006 \\
IIZw70-SW 	&3x1200 	&V1200 	&3610-4379 &04/06/2006 \\

\hline
\end{tabular}
\end{table}

\subsection{Data reduction}

We reduced the data using the software R3D (S\'anchez 2006). Different
exposures taken at the same pointing were combined using
IRAF\footnote{IRAF is distributed by the National Optical Astronomy
Observatories.} tasks.  The expected locations of the spectra were
traced on a continuum-lamp exposure obtained before each target
exposure. After bias subtraction, we extracted the target spectra by
adding the signal from the 5 pixels around the central traced pixel
(that is the total object spectrum width).  With exposures of Hg and
Ne lamps obtained immediately after the science exposures, the spectra
were wavelength calibrated. We checked the accuracy of the wavelength
calibration using sky emission lines, and find standard deviations of
0.5 \AA~ and 0.3 \AA~ for the V300 and V1200 gratings, respectively.
The continuum-lamp exposure was also used to determine the response of
the instrument for each fiber and wavelength (the so-called
flat-spectra). We used these flat-spectra in order to homogenize the
response of all the fibers. The average nearby background spectrum,
obtained moving the IFU off-target, was subtracted from the target
spectra. For the standard star observations we co-added the spectra of
the central fibers and compared the one-dimensional standard star
spectrum with table values to create a sensitivity function. The
spectra were flux calibrated using IRAF.

The reduced spectra were contained in two data cubes, each one
corresponding to a different PMAS pointing.  The two data cubes were
mosaiced into a single frame (see the blue box mosaic in
figure~\ref{SDSS}). We corrected the mosaic data cubes for the effect
of differential atmospheric refraction using the R3D package
(S\'anchez 2006). We recentred and scaled the mosaic data cube taken
in 2006 with respect to the mosaic data cubes obtained in 2005 making
use of the H$\gamma$ emission line, common to the wavelength ranges of
both gratings. Further manipulation of the cubes, such as the
production of emission-line maps was performed using E3D (S\'anchez
2004).

\subsection{Line intensities}

Using the software FIT3D (S\'anchez et al. 2007), we fitted line
profiles on the extracted one-dimensional spectra in order to derive
the integrated flux of each emission line.  A single gaussian was
fitted to each emission line, using a polynomial function to
characterize the continuum. The software FIT3D allows definition of
emission line systems, i.e., a kinematically coupled set of emission
lines with the same width. This was essential for accurate deblending
of the lines, when necessary.

\begin{figure}[!ht]
 \mbox{
  \centerline{
\hspace*{0.0cm}\subfigure{\label{}\includegraphics[width=7cm]{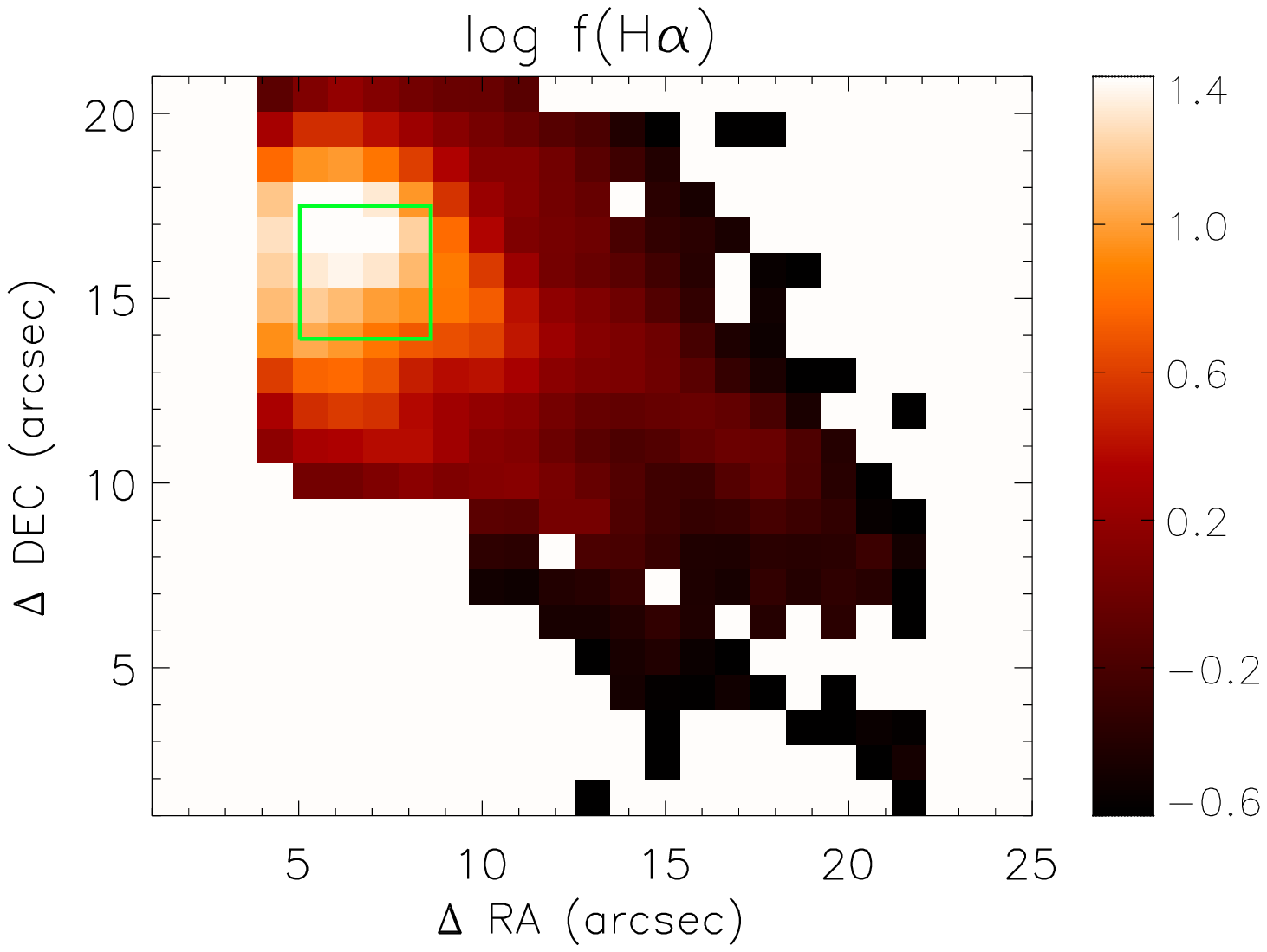}}
   }}
 \mbox{
  \centerline{
\hspace*{0.0cm}\subfigure{\label{}\includegraphics[width=7cm]{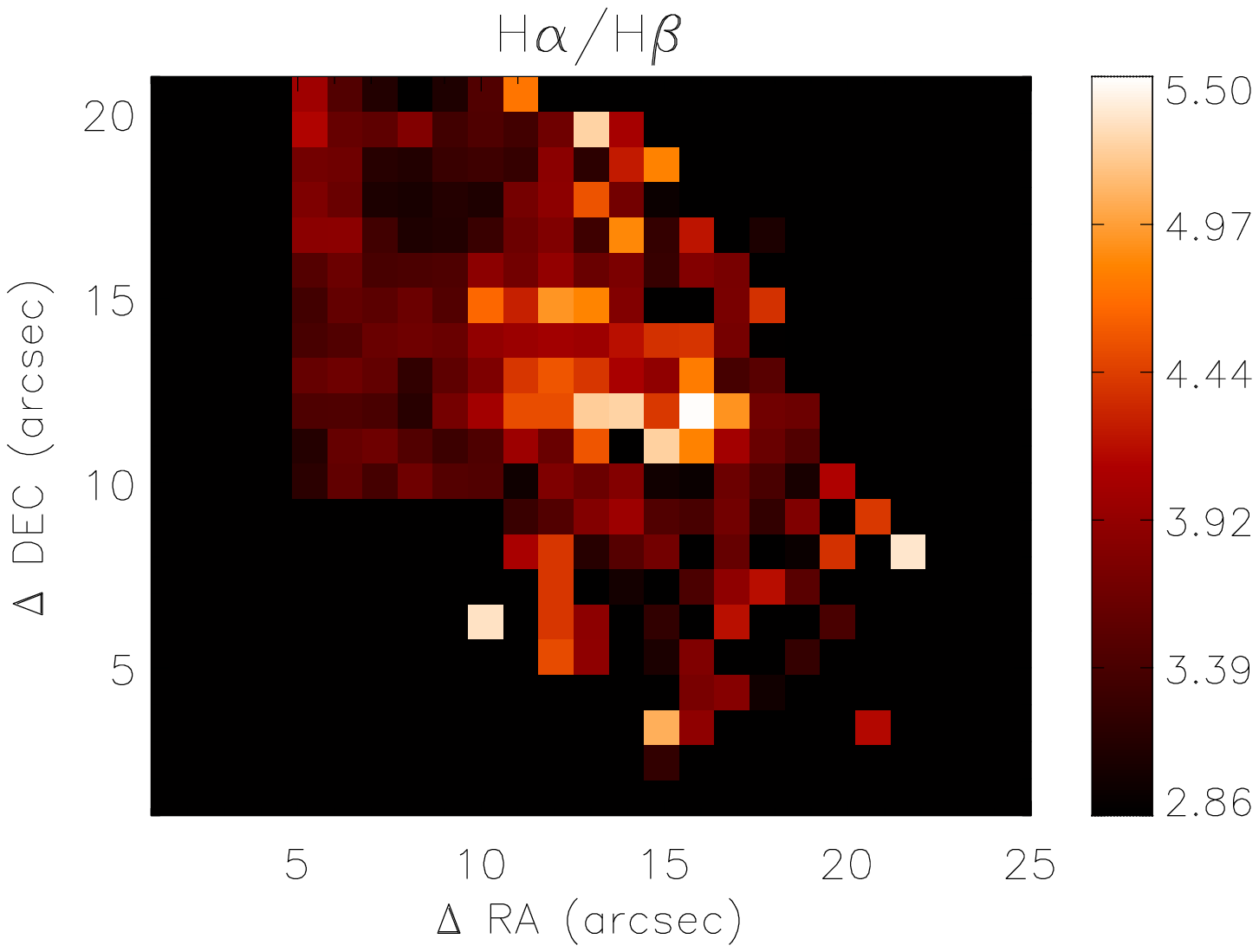}}
   }}
  \caption{Top image: map of the log flux(H$\alpha$) in each pixel in units of 10$^{-15}$ erg s$^{-1}$ cm$^{-2}$; H$\alpha$ fluxes with  relative errors $\geq$ 10$\%$ are not shown. We overplot the exact location of the green box from figure 1 for orientation. Bottom image: H$\alpha$/H$\beta$ ratio spatial distribution. The lowest values tend to be concentrated near to the brightest region of the galaxy. Only H$\alpha$/H$\beta$ ratios with relative errors $\leq$ 30$\%$ are shown. North is up, east is left. Relative right ascension (RA) and declination (DEC) are shown in arcsec.}
\label{HA}
\end{figure}

For each fiber spectrum we derived its corresponding reddening
coefficient, C(H$\beta$), using the value of the Balmer decrement
derived from H$\alpha$/H$\beta$, as compared to the theoretical value
expected for case B recombination from Storey $\&$ Hummer (1995), and
applying the extinction law given by Whitford (1958). Thus, the fluxes
of the emission lines for each fiber were corrected for extinction
using their corresponding C(H$\beta$) value. 

Equivalent widths (EWs) of Balmer emission lines are so high
[e.g. EW(H$\beta$) $\sim$ 40-120 \AA] that the effect of the
underlying stellar population in these lines appears not to be
important [underlying absorption-lines EWs present typical values of
1-2 \AA~; see e.g. McCall et al. (1985)].  In addition, we carried out
an eye-inspection, fiber-by-fiber, of the H$\beta$ and H$\gamma$
emission lines, and we did not find any stellar absorption features
(e.g. wings of absorption lines) underlying these emission lines.

We calculated the error in the line fluxes, $\sigma_{l}$, from the
expression $\sigma_{l}$ =
$\sigma_{c}$N$^{1/2}$[1+EW/N$\Delta$]$^{1/2}$ (Catellanos 2000) where
$\sigma_{c}$ represents a standard deviation in a box centred close to
the measured emission line, N is the number of pixels used in the
measurement of the line flux, EW is the equivalent width of the line,
and $\Delta$ is the wavelength dispersion in \AA/pixel. This
expression takes into account the error in the continuum and the
photon counts statistics of the emission line. The error
measurements were performed on the extracted one-dimensional spectra.

H$\alpha$ emission line map (continuum subtracted) and the spatial
distribution of H$\alpha$/H$\beta$ are shown in figure~\ref{HA}. The
dust distribution seems to be non uniform across the galaxy. We would
like to stress that in the following sections, all emission line
fluxes, used to derive the ionization structure and physical-chemical
parameters, are corrected for extinction using the corresponding value
of C(H$\beta$) for each fiber.

\section{Results}

\subsection{Galaxy structure}


\begin{figure}
 \mbox{
  \centerline{
\hspace*{0.0cm}\subfigure{\label{}\includegraphics[width=4cm]{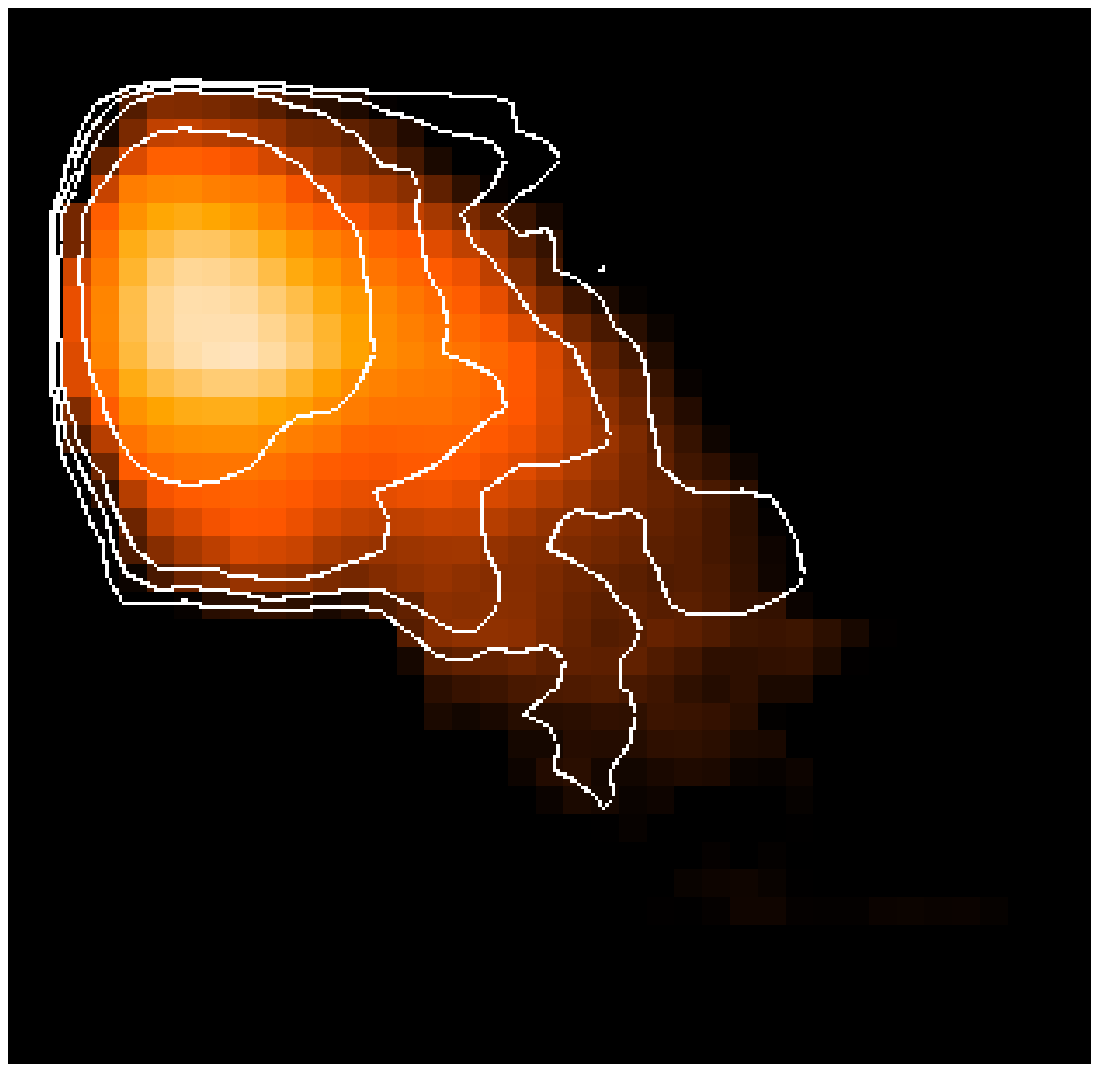}}
\hspace*{0.0cm}\subfigure{\label{}\includegraphics[width=4cm]{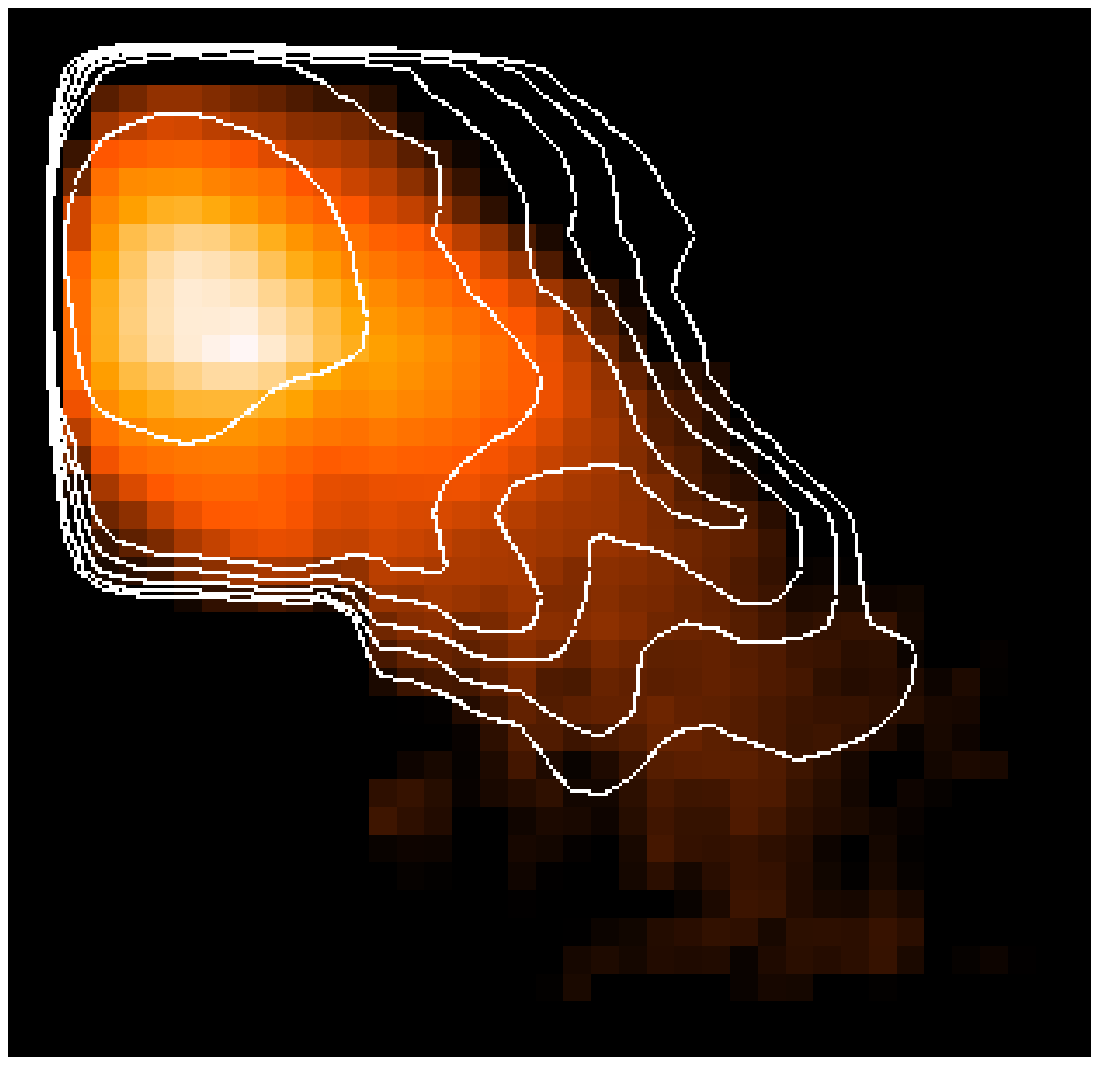}}
   }}
  \caption{Maps in the continuum near H$\beta$ ({\it left}) and H$\alpha$ ({\it
right}) emission lines; isocontours of the corresponding emission line
fluxes are shown overplotted. Maps of the continua are representative
of the stellar emission, free of contamination from the gaseous
emission lines. Isocontours and maps are displayed in logarithmic
scale, with north up and east to the left. The image size is $\sim$
$24\hbox{$^{\prime\prime}$ }\times23\hbox {$^{\prime\prime }$}$. The
pixel size is $\sim$ 0.6\hbox{$^{\prime\prime }$}.}

\label{maps2}
\end{figure}

\begin{figure}
\center{
\includegraphics[width=5cm,clip]{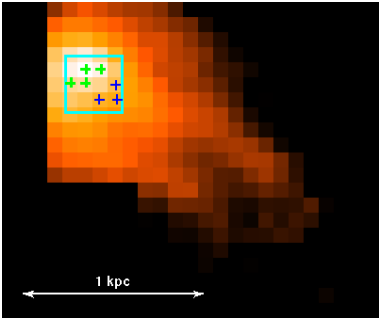}}
\caption{The zone occupied by the probable brightest clusters (blue rectangular box with  physical dimension as green box in figure~\ref{SDSS} ({\it right})), detections of the WR bump (green crosses) and nebular HeII$\lambda$4686 emission (blue crosses) on H$\alpha$ flux map  (see text for details). North is up, east is left. The size of the image is  
$25\hbox{$^{\prime\prime}$ }\times21\hbox {$^{\prime\prime }$}$ and the
pixel size is 1\hbox{$^{\prime\prime }$}. The spatial scale is shown, where 1 kpc $\sim$ 11\hbox{$^{\prime\prime }$} assuming a distance to the galaxy of 18.1 Mpc (see table 1).}
\label{cumulo.WR.HeII}
\end{figure}

In figure~\ref{maps2} we show the continuum emission near H$\beta$ and
H$\alpha$ maps, together with the corresponding H$\beta$ and H$\alpha$
emission line intensity contours overplotted. The galaxy's morphology
appears very similar in emission lines and in continuum emission,
i.e. the continuum and emission line maps (as seen from the contours
in figure~\ref{maps2}) show a central region with ionized gas and a
faint extension towards the south-west. However we note that in the
emission line contours we can resolve clearly two filamentary
structures, with a tail-like shape, at the faintest extensions of the
galaxy which are not present in the continuum maps.

The {\it r} filter image in figure~\ref{SDSS} shows a red extension
already found by Cair\'os (2001) and Kong et al. (2003). Such a red
extension could indicate the likely existence of an intermediate-old
underlying stellar population. Kong et al. (2003) studied the stellar
content of a sample of 73 HII galaxies by analysing their continuum
spectra and absorption features. They showed that in most of HII
galaxies (the galaxy IIZw70 is among them), stars older than 1 Gyr
exist. The average colours of the low-surface-brightness component
(B-V = 0.47, V-I = 0.89) are also consistent with an evolved stellar
background with an age of 1-3 Gyr for IIZw70, as shown by Cair\'os
(2001).

The {\it u} band (right image in figure~\ref{SDSS}) is expected to
trace the youngest stellar populations. The box in this image
coincides with the {\it u} band maximum and encloses the region
occupied by the probable brightest young clusters.  We also show the
exact location of this box in the {\it r} image of IIZw70 as a
reference (left image). We verified that the location of the probable
youngest stellar population, as seen in the {\it u} image, coincides
with the central bright SF knot and with the zone of maximum
in the continuum emission (see figures~\ref{maps2} and
~\ref{cumulo.WR.HeII}).

\subsection{Ionization structure and excitation sources}

\begin{figure}[!ht]
 \mbox{
  \centerline{
\hspace*{-0.2cm}\subfigure{\label{}\includegraphics[width=5.3cm]{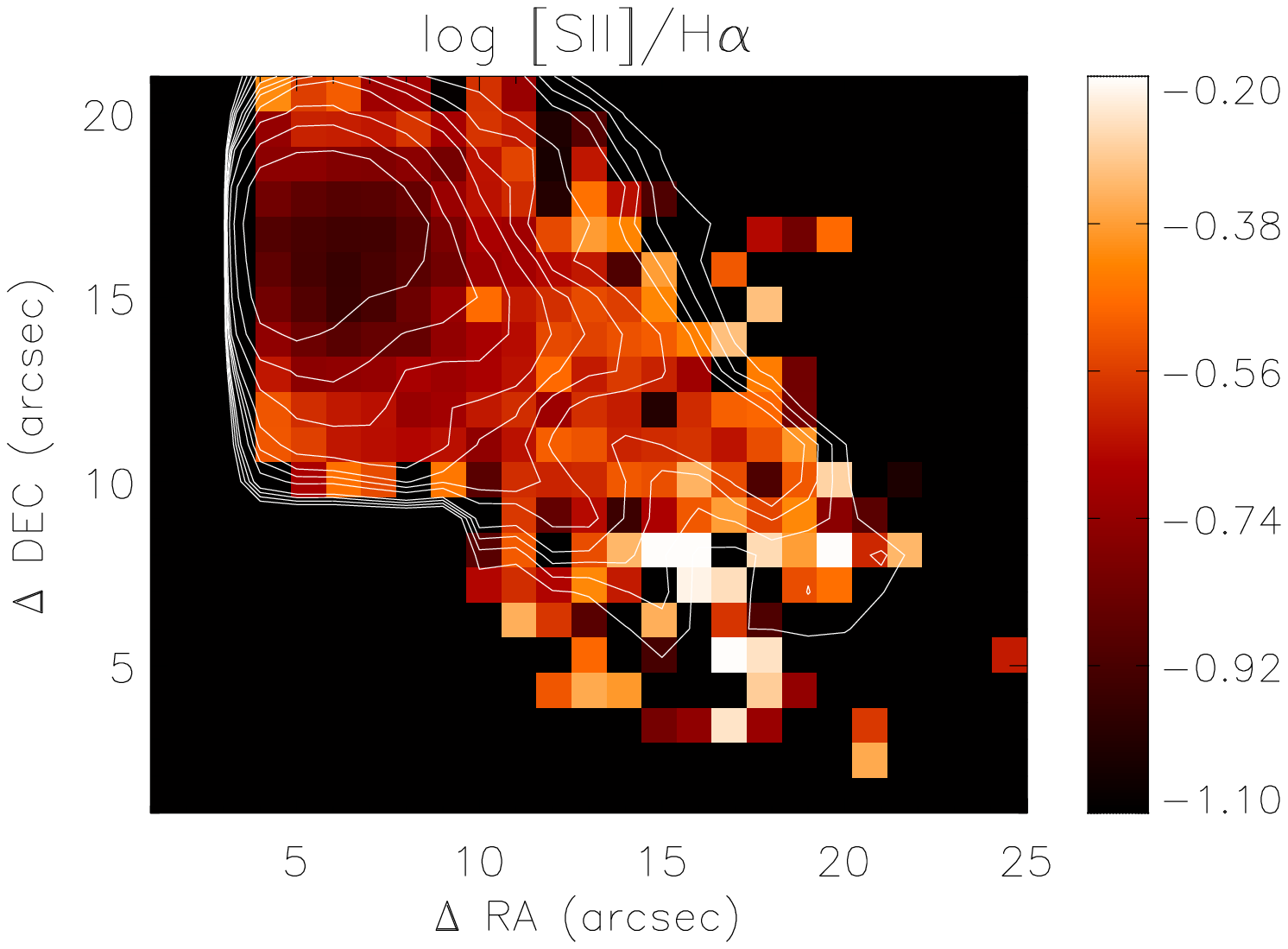}}
\hspace*{-1.0cm}\subfigure{\label{}\includegraphics[width=5.3cm]{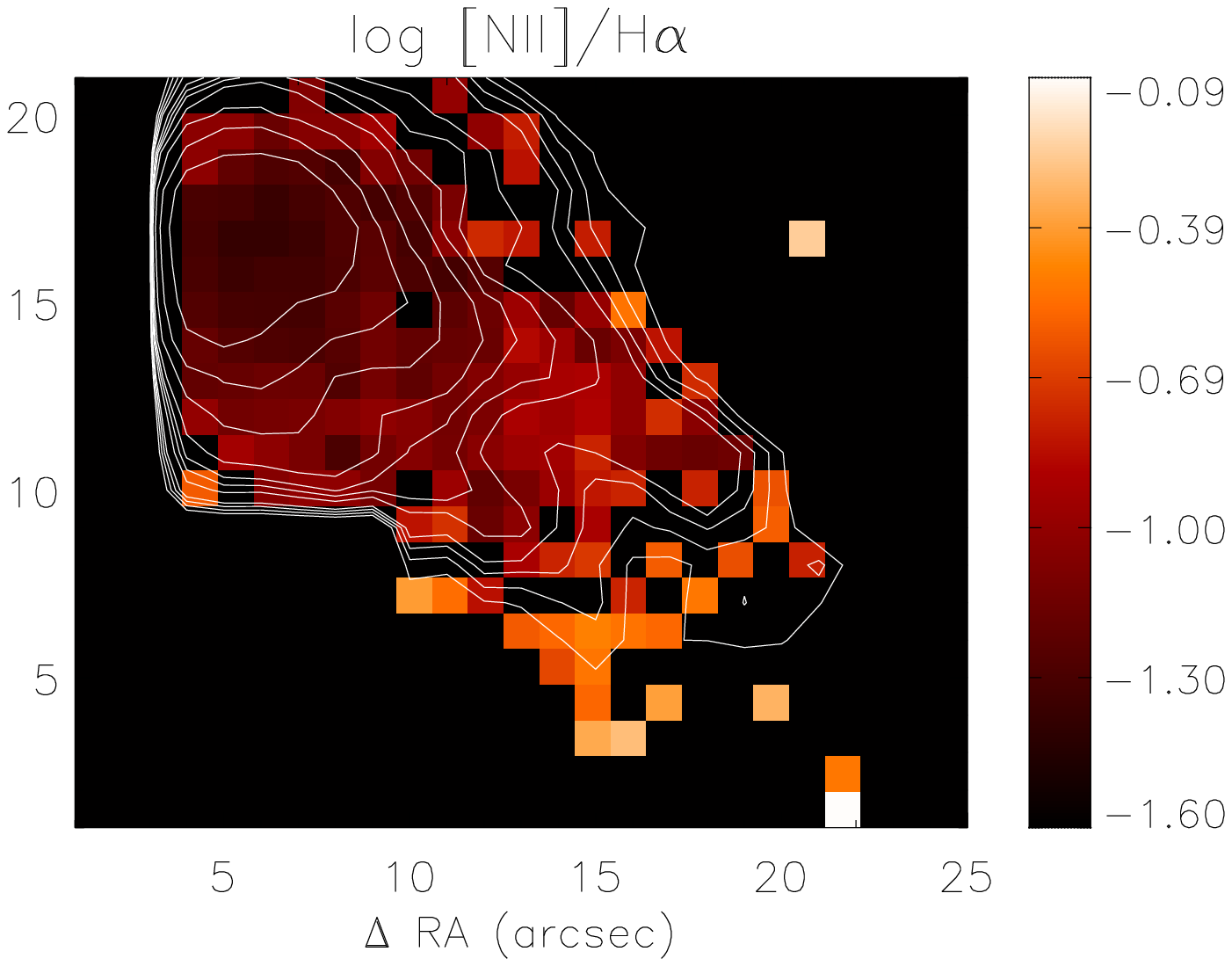}}
   }}
 \mbox{
  \centerline{
\hspace*{0.0cm}\subfigure{\label{}\includegraphics[width=5.3cm]{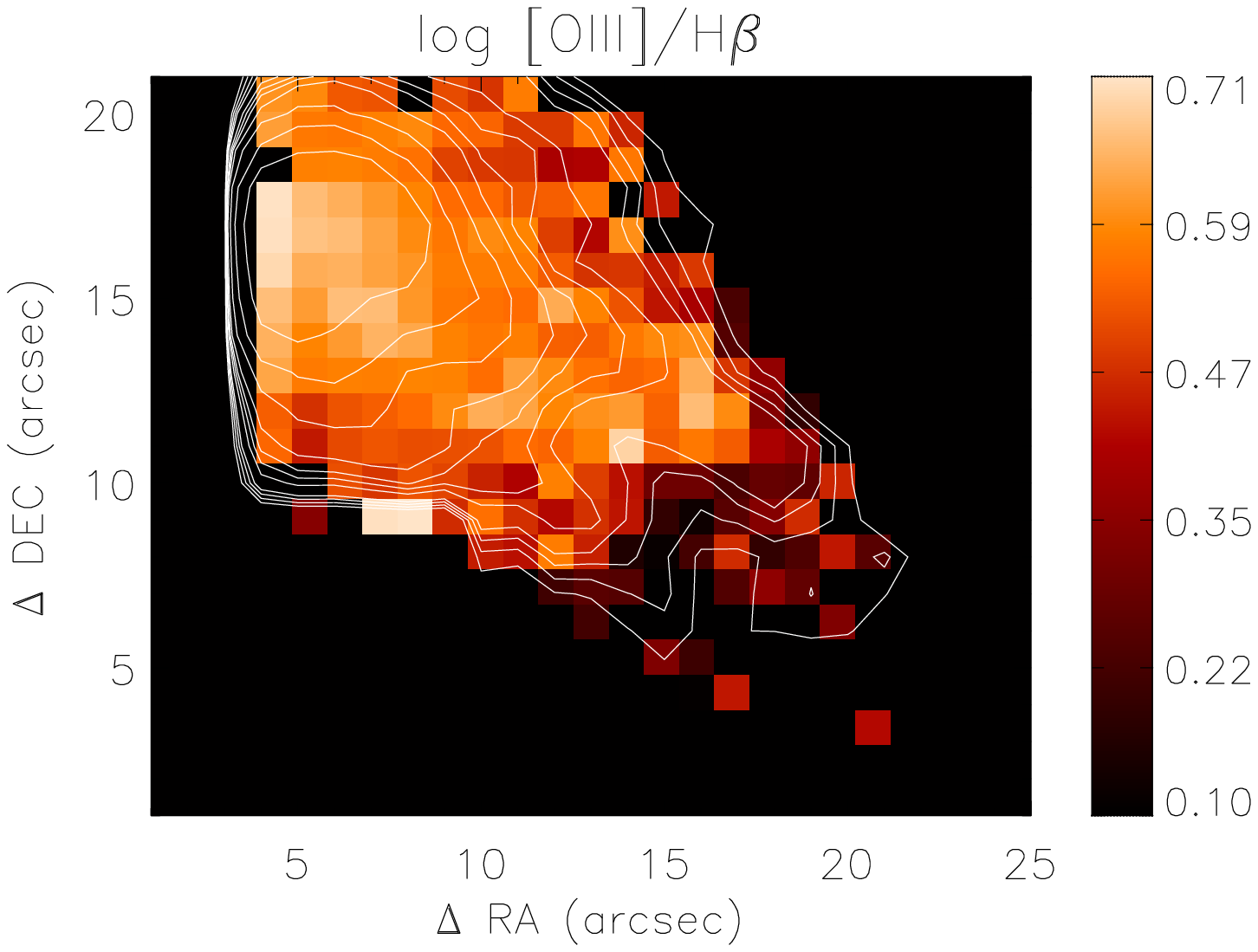}}\\
   }}
  \caption{Emission line ratio maps in logarithmic scale:
[SII]$\lambda$6717.31/H$\alpha$ (top left panel),
[NII]$\lambda$6584/H$\alpha$ (top right panel) and
[OIII]$\lambda$5007/H$\beta$ (bottom panel). Only line-intensity
ratios with relative errors $\leq$ 30$\%$ are represented. Isocontours of the
H$\alpha$ emission line flux are shown overplotted as reference. North is
up, east is left. Relative right ascension (RA) and declination (DEC)
are shown in arcsec.}

\label{diag1}
\end{figure}

\begin{figure}
   \includegraphics[bb=0 255 570 520,width=\columnwidth,clip]{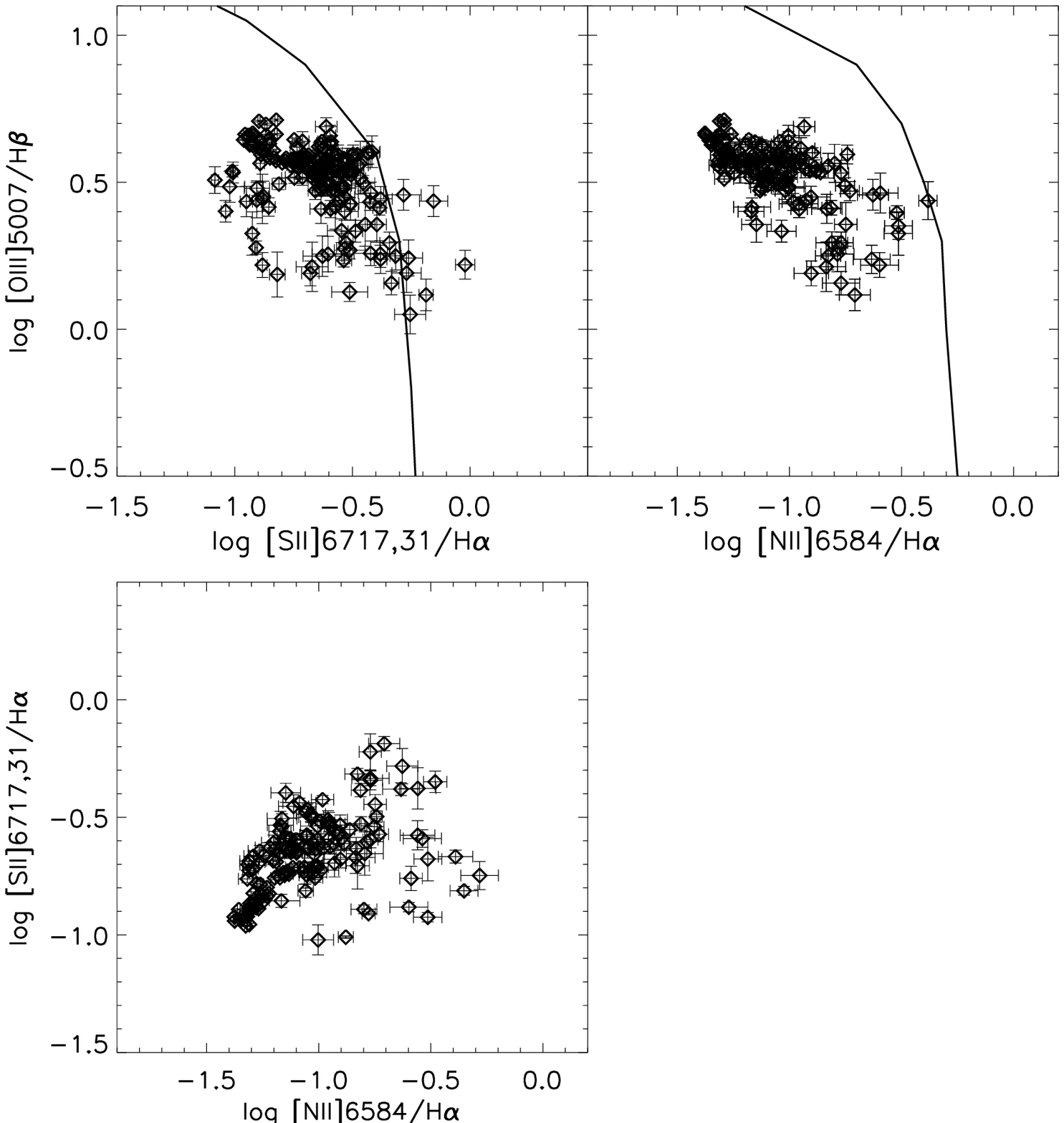}
      \caption{Left panel shows the relation between
      [OIII]$\lambda$5007/H$\beta$ and [SII]$\lambda$6717.31/H$\alpha$, and right panel
      presents the relation between [OIII]$\lambda$5007/H$\beta$ and
      [NII]$\lambda$6584/H$\alpha$ in each fiber. The solid lines are adapted from Osterbrock \& Ferland
(2006) (see the text for details).}
         \label{diag2}
   \end{figure}

The line ratio maps for the usual diagnostic lines (e.g.  Baldwin et
al. 1981; Osterbrock \& Ferland 2006) can be used to investigate the
spatial distribution of the ionization structure of the ionized
nebulae. Figure~\ref{diag1} shows the line ratio maps (in logarithmic
scale) for
[NII]$\lambda$6584/H$\alpha$,[SII]$\lambda$6717,31/H$\alpha$,
[OIII]$\lambda$5007/H$\beta$. H$\alpha$ emission line flux contours
are plotted over all maps. High excitation values correspond to low
values of the first two ratios and to high values of the last one.

It can be seen that [NII]$\lambda$6584/H$\alpha$ as well as
[SII]$\lambda$6717,31/H$\alpha$ increases in zones where the H$\alpha$
intensity decreases; [OIII]$\lambda$5007/H$\beta$ presents the
opposite trend. This behaviour towards lower excitation when the
intensity is smaller can be related to the increasing distance from
the young stellar clusters, with the corresponding decrease in the
local values of the ionization parameter, as it is typical in extended
HII regions (see McCall et al. 1985)

Figure~\ref{diag2} shows the classical BPT diagnostic diagrams
(Baldwin et al. 1981) using our own data. Only fibers that present
diagnostic line-intensity ratios with relative errors $\leq$ 30$\%$
are plotted.  The solid lines (adapted from Osterbrock \& Ferland 2006) show
the locus of separation between regions dominated by
photoionization (HII-regions like; left of the line) and
regions dominated by shocks.  Our data show that the line ratios for
most positions in the galaxy are located in the general locus of
HII-region like objects. 

There are a few fibers, located at the tail-like regions (see the
[SII]$\lambda$6717,31/H$\alpha$ map in figure~\ref{diag1}), with
values of [SII]$\lambda$6717,31/H$\alpha$ somewhat higher that are
found outside the HII-region zone. These values are consistent with
predictions from shock models (Dopita \& Sutherland 1995) suggesting
the presence of shocks in the ionized gas outflow scenario proposed in
section 3.5. Nonetheless, we must keep in mind that the line ratios of
both axis plotted in figure~\ref{diag2} are sensitive to the
metallicity, as well as, to the ionization parameter. Besides, as
these line ratios can vary within an HII galaxy (i.e. across giant HII
regions), thus BPT diagnostic diagrams on spatially resolved data of
HII galaxies should be studied in depth though that is not the purpose
here.

\subsection{Wolf-Rayet stellar population}

WR stars are the evolved
descendants of massive O progenitor stars (Maeder \& Conti
1994). Stellar evolution theory predicts few WR stars to form in a
low-metallicity environment such as in IIZw70 (e.g. Meynet
1995). Therefore WR star detections in low metallicity HII galaxies
have important implications for the stellar evolution theory. A 
common property of HII galaxies with WR stars (WR galaxies), provided
that our understanding of stellar evolution is correct, is the existence of ongoing
active or recent SF which has produced stars massive
enough to evolve to the WR stage. This typically indicates  ages of a few
Myr for stars with initial masses $M_{ini}$ $\gtrsim$ 20$M_{\odot}$
(Maeder \& Conti 1994).

The presence of WR stars can be recognized via the WR bumps around
$\lambda$4650 $\AA$~(blue bump) and $\lambda$5808 $\AA$~(red bump), which are generally a
blend of HeII and several metal lines. The blending of several
stellar and nebular emission lines around $\lambda$4686 $\AA$ and the complex
spatial morphology in ground-based data make it challenging to
disentangle the emission of stars and gas and to derive the WR content
and the corresponding nebular properties.

\begin{figure}
\center{
\includegraphics[width=7cm,clip]{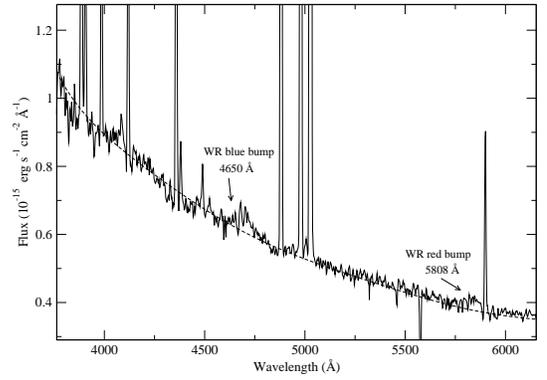}}
\caption{Spectrum showing the WR bumps obtained adding the emission from the 4 fibers ($2\hbox{$^{\prime\prime}$ }\times2\hbox {$^{\prime\prime }$}$) marked with a green cross in figure~\ref{cumulo.WR.HeII}. Dashed line is the continuum fit.}
\label{BUMP}
\end{figure}

For the first time we detected WR stars in
IIZw70. Summing the emission from 4 fibers (green crosses in
figure~\ref{cumulo.WR.HeII}) we can observe the blue and red bumps, --
the latter marginally detected --, in the corresponding spectrum (see
figure~\ref{BUMP}). This detection would indicate that WR bumps are
being seen in an extended area ($\sim$ 4 arcsec$^{2}$). The use of the
IFS minimises the difficulty in correlating spectra and image
location, shown in other works (e.g. de Mello et al. 1998).  The WR
bumps intensity maximum coincides with the likely youngest clusters
location (as shown in the {\it u} SDSS image; see figures~\ref{SDSS}
and \ref{cumulo.WR.HeII}) and, within 1$^{\prime\prime}$, also
spatially coincides with the peak of the H$\alpha$ emission. In
addition we verified that the maximum of the continuum emission
adjacent to H$\beta$ ($\sim$ $\lambda$4800 \AA) corresponds spatially
to the WR bumps intensity maximum.  From the spectrum shown in
figure~\ref{BUMP} we measured a value of EW(WRbump) of -2.2 $\pm$ 0.6
\AA~ for the WR blue bump and a corresponding 100 x
I(WRbump)/I(H$\beta$) of 4.6 $\pm$ 1.1 .

An area of HeII$\lambda$4686 narrow emission was found in 3 fibers
located in the southwest corner of the box encompassing the probable
central brightest clusters (see figure~\ref{cumulo.WR.HeII}); the line
is narrow showing approximately the same width as the [OIII] lines,
confirming its correct identification as a nebular line. The presence
of the nebular HeII$\lambda$4686 line should indicate the existence of
a very hard ionizing spectrum.

\subsection{Physical properties and abundance analysis}
\label{total_abund}

We calculated the physical properties and ionic abundances of the
ionized gas for IIZw70 following the 5-level atom FIVEL program (Shaw
\& Dufour 1994) available in the task IONIC of the STSDAS
package. Total abundances were derived as we detail in the
following. We calculated the final quoted errors in the derived
quantities by error propagation taking into account errors in flux
measurements.

We obtained the electron densities (N$_{e}$) from the
[SII]$\lambda$6717/$\lambda$6731 line ratio.  Although
[OII]$\lambda$3726,29 emission lines are brighter than [SII]
$\lambda$6717,31, the spectral resolution is at the limit to resolve
accurately the [OII] doublet lines. [OII] densities are roughly
consistent with [SII] densities within the errors.

\begin{figure}
\center{
\includegraphics[width=5cm,clip]{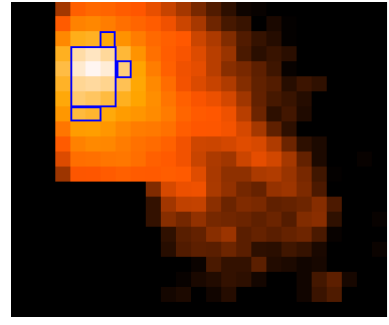}}
\caption{[OIII]$\lambda$5007 flux map. The blue box marks the zone of the galaxy where we could measure the [OIII]$\lambda$4363 emission line. The size of the image is  
$25\hbox{$^{\prime\prime}$ }\times21\hbox {$^{\prime\prime }$}$ with a
pixel size of 1\hbox{$^{\prime\prime }$}. North is up, east is left.}
\label{OIII.ZONAS.ABUND}
\end{figure}

We derived the electron temperature values of T$_{e}$[OIII] using the
[OIII]$\lambda$4363/$\lambda$4959,5007 line ratio.  T$_{e}$[OII] was
calculated from the relation between [OII] and [OIII] electron
temperatures given by Pilyugin et al. (2006). 
We were able to measure the faint auroral line [OIII]$\lambda$4363 for
16 fibers with high S/N leading to relative errors lower than 10$\%$. These fibers
cover a projected area of nearly 19 arcsec$^{2}$ equivalent to
$\sim$ 0.4 $\times$ 0.3 kpc$^{2}$, including the central starburst
region. This area is shown in figure~\ref{OIII.ZONAS.ABUND} superposed
on the H$\alpha$ emission map. T$_{e}$[OIII] values used to derive the
ionic abundance O$^{++}$ were derived from emission line fluxes from
the high spectral resolution observations in order to avoid any
possible contamination of $\lambda$4363 by terrestrial HgI
emission. Nevertheless we compared the T$_{e}$[OIII] values derived
from both our low and high spectral resolution observations, finding
that the measurements coincide within the errors.

\begin{figure}[ht]
 \mbox{
  \centerline{
\hspace*{-0.2cm}\subfigure{\label{temp_zone}\includegraphics[bb=130 683 247 853,width=5cm,clip]{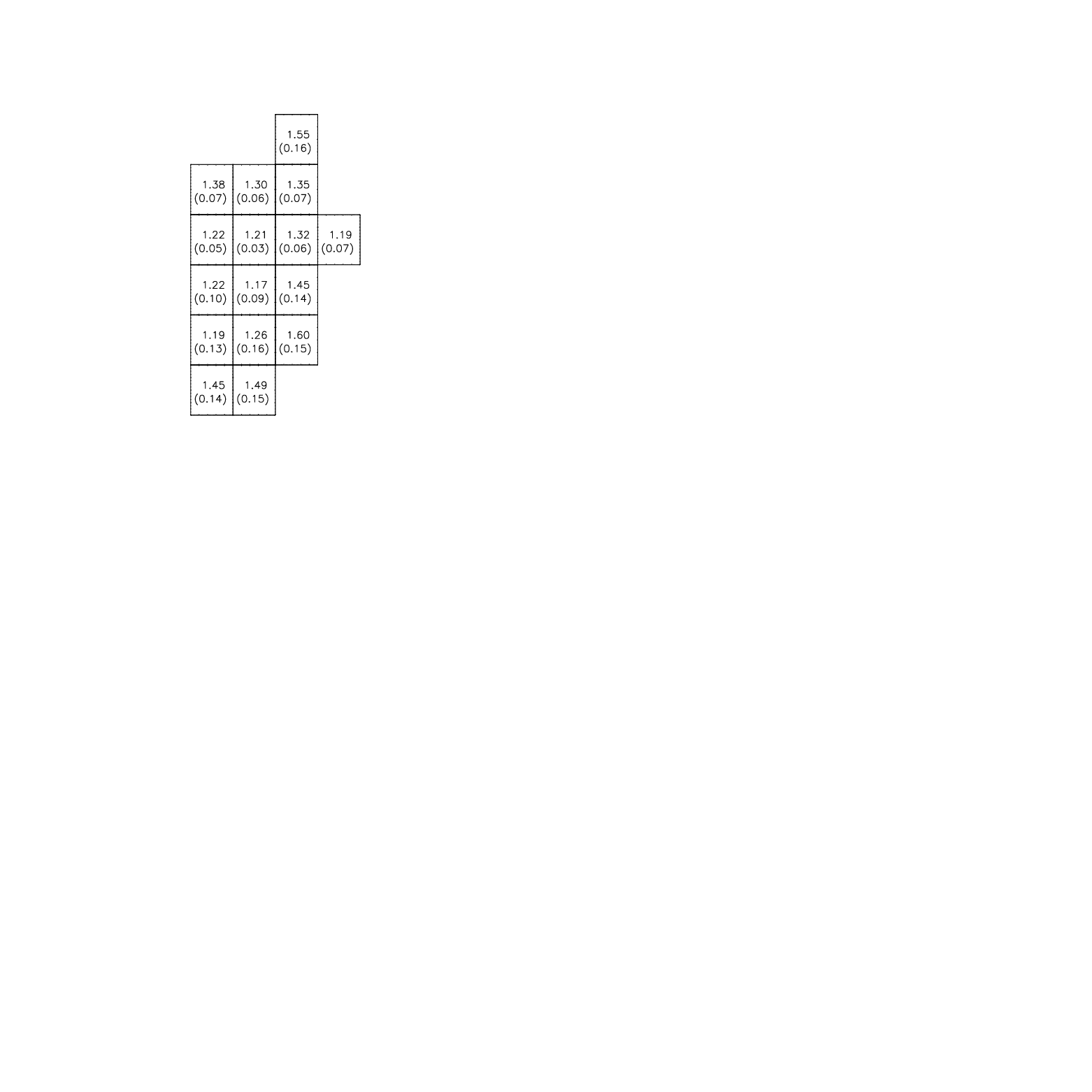}}
\hspace*{-1.0cm}\subfigure{\label{dens_zone}\includegraphics[bb=130 683 247 853,width=5cm,clip]{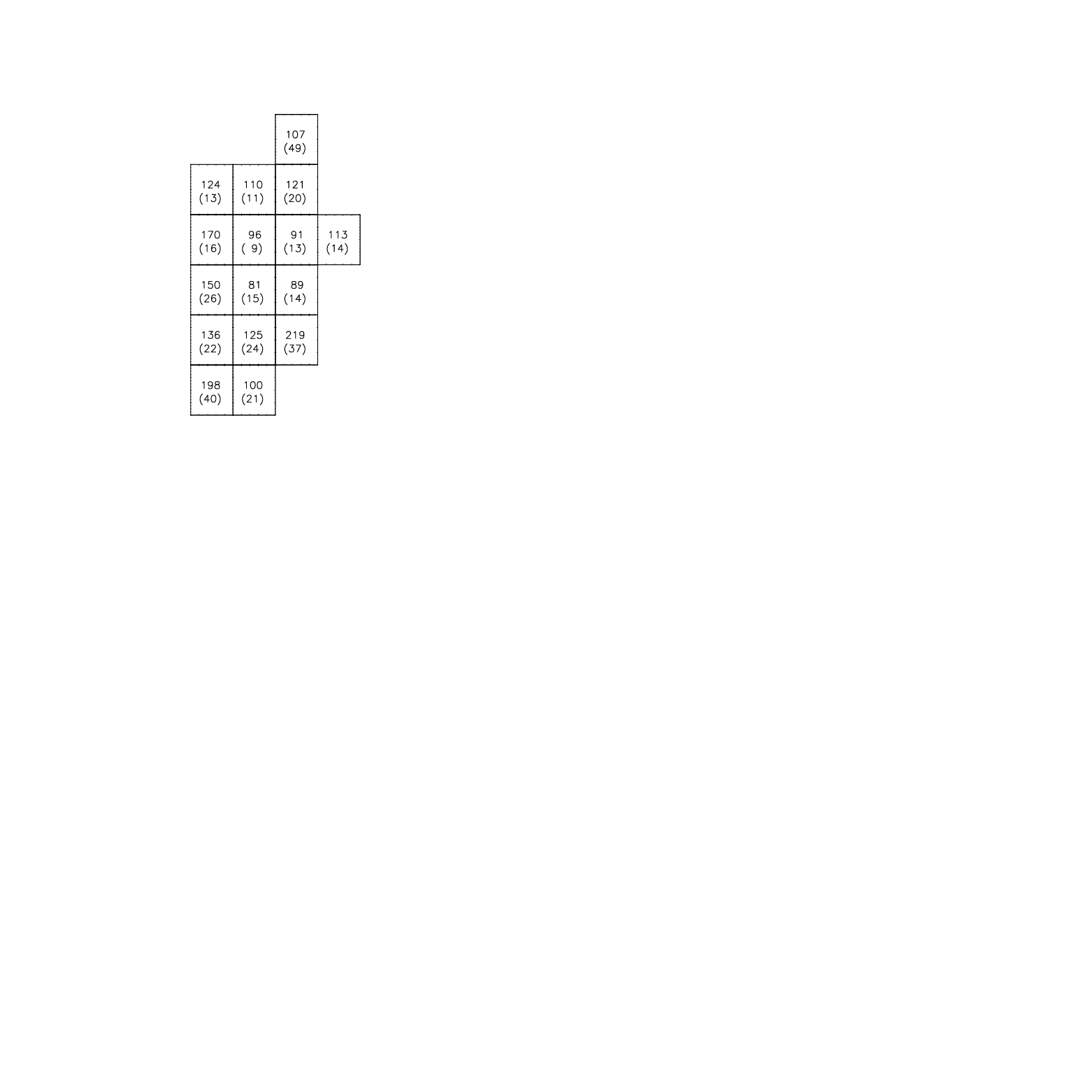}}
   }}
 \mbox{
  \centerline{
\hspace*{0.0cm}\subfigure{\label{abund_zone}\includegraphics[bb=130 683 247 853,width=5cm,clip]{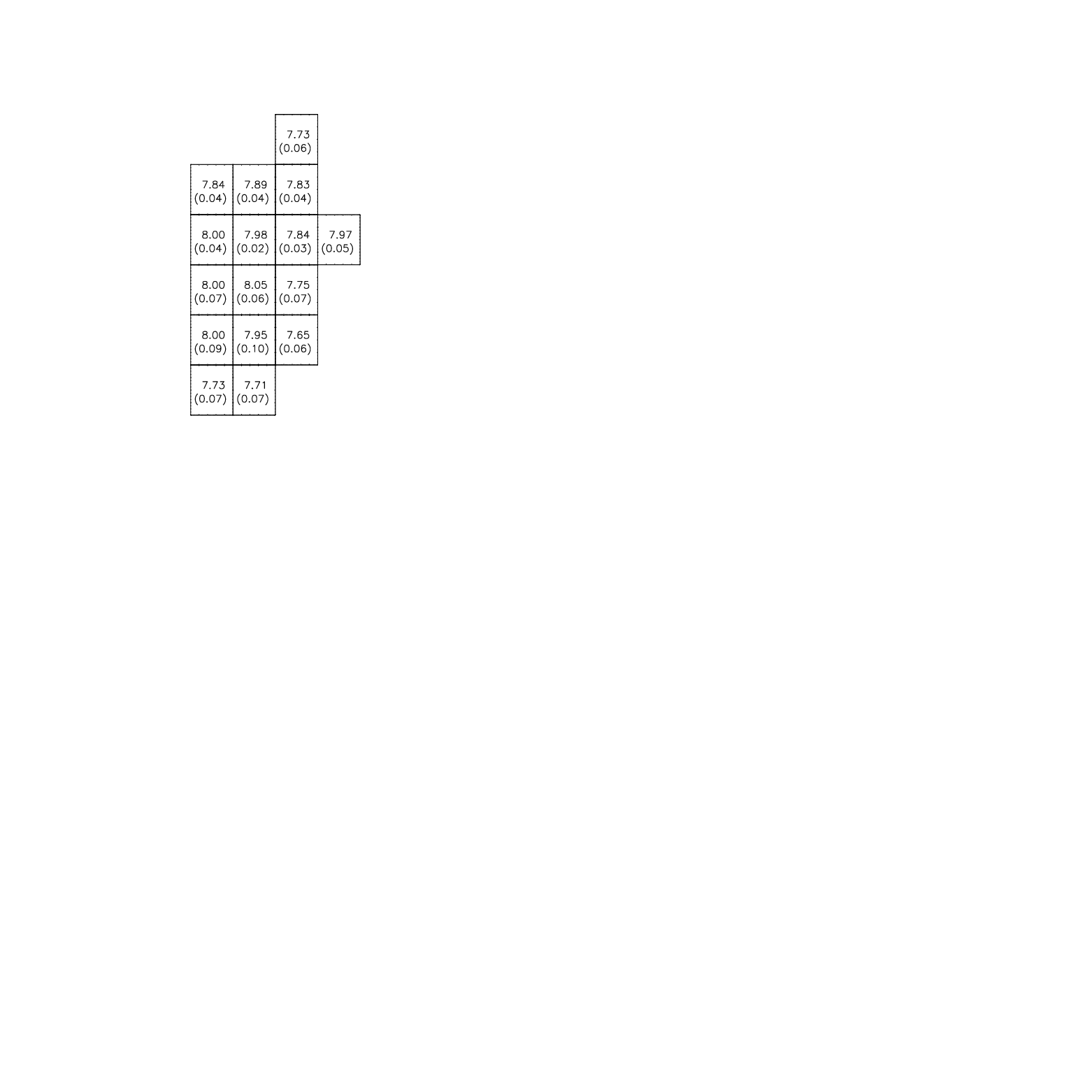}}\\
   }}
  \caption{Top left panel: electron temperature distribution (in 10$^{4}$ K) from the [OIII]
$\lambda$4363/($\lambda $4959+$\lambda $5007) line ratio. Top right panel: electron
density distribution (in cm$^{-3}$) from the
[SII]$\lambda$6717/$\lambda$6731 line ratio. Bottom panel: the distribution of
the derived oxygen abundance, 12+log O/H. Note that the regions shown
are outlined by the blue box in figure~\ref{OIII.ZONAS.ABUND}.}

\label{cond_fis}
\end{figure}

Figure~\ref{cond_fis} shows the distribution of the electron
temperature T$_{e}$[OIII] (top left panel), electron number density
derived from the [SII]$\lambda$6717,31 emission lines (top right
panel) and the total oxygen abundance spatial distribution (bottom
panel).  The errors are quoted in parenthesis. We show only the fibers
where the [OIII]$\lambda$4363 emission line could be well measured.

We did not find any statistically significant variance for the
electron number density; N$_{e}$ values are of the order 100 cm$^{-3}$
approximately, within the errors. We saw that there are some
variations in the derived electron temperature (11700 - 16000 K). The
derived values of T$_{e}$[OIII] versus the error in the
[OIII]$\lambda$4363 line were plotted and no correlation was found,
indicating that no significant noise contribution near $\lambda$4363
was included in the flux measurements.

With regards to the oxygen ionization correction factor (ICF), a small
fraction of O/H is expected to be in the form of the O$^{3+}$ ion in
the high-excitation HII regions when the HeII$\lambda$4686 emission
line is detected. We have a measurement of the HeII$\lambda$4686
emission line in 3 fibers. According to the photoionization models
from Stasi\'nska \& Izotov (2003), the O$^{3+}$/O can be in the order
of 1$\%$ only in the highest-excitation HII regions [O$^{+}$/(O$^{+}$
+ O$^{2+}$) $\leq$ 0.1]; therefore, taking our abundance results into
account, this ICF correction was assumed to be negligible in our case.

The oxygen abundance derived for the 16 positions studied does not
show correlation with the ionization degree over the whole range in
O$^{2+}$/O$^{+}$. There are some variations in the derived oxygen
abundance in the range 12+log(O/H)= 7.65 to 8.05.  The error-weighted
mean of the derived oxygen abundance values is 12 + log(O/H) = 7.86
$\pm$ 0.05 (1$\sigma$). 14 out of 16 O/H measurements are consistent
with this mean value at 2$\sigma$ ($\pm$ 0.1 dex).  The other 2
measurements present values of O/H consistent, within the errors, with
$\pm$ 3$\sigma$ of the weighted mean. Bearing in mind that the errors
quoted in the oxygen abundance do not account for some observational
uncertainties (e.g. pointing errors, seeing variations, etc) nor for
errors associated to the reddening correction and flux calibration, we
must assume that the variations in the derived oxygen abundance may
not be statistically significant.

In our spectra, neon is seen via the [NeIII] emission line at
$\lambda$3869~\AA.  For this ion, we took the [OIII] electron
temperature as representative of the high excitation zone. We
calculated the neon abundance assuming that

\[Ne/O = Ne^{2+}/O^{2+}\]

Regarding [SII] and [NII] temperatures we assumed the approximation
T$_{e}$[SII] $\approx$ T$_{e}$[NII] $\approx$ T$_{e}$[OII] as valid, since no 
auroral line could be measured in the low excitation zone.
The N/O abundance was derived under the assumption that \[N/O=N^{+}/O^{+}.\]


Regarding N/O and Ne/O abundance ratios, evaluating the contribution
of all observational errors to the derivation of these abundances, it
can be concluded that there is no statistical evidence for any
systematic variation of N/O and Ne/O [-1.15$\pm$0.09 $\leq$ log(N/O)
$\leq$ -0.90$\pm$0.11; -1.07$\pm$0.18 $\leq$ log(Ne/O) $\leq$
-0.84$\pm$0.10] with O/H for the range of abundances studied in this
work.  The values of the error-weighted mean for log(N/O) and for
log(Ne/O), -1.04 $\pm$ 0.06 dex and \mbox{-0.97 $\pm$ 0.08}
respectively, are consistent, within the errors, with the log(N/O) and
log(Ne/O) solar values, -0.86 $\pm$ 0.12 and \mbox{-0.82 $\pm$ 0.11}
(Lodders 2003), with N/O being slightly ($\sim$ 0.1 dex) lower than
solar.  The error-weighted mean value for log(N/O) in IIZw70 is
significantly higher ($\sim$ 0.4 dex) than the mean for metal-poor HII
galaxies [log(N/O) = -1.46 $\pm$ 0.14 for galaxies with 7.6 $<$
12+log(O/H) $<$ 8.2; Izotov \& Thuan 1999]. Still the N/O ratio of
IIZw70 is consistent with values found for a few HII galaxies in the
literature (Nava et al. 2006, Izotov \& Thuan 1999) with similar
metallicity. As regards Ne/O abundance ratio, the error-weighted mean
value for log(Ne/O) derived here (-0.97 $\pm$ 0.08) is lower than the
mean value of -0.72 $\pm$ 0.07 obtained by Izotov \& Thuan
1999. However, they also found some galaxies with log(Ne/O) similar to
the values derived in this paper. In any case we should bear in mind
that neither the literature long-slit measurements we are comparing to
nor our IFU mosaic cover the entire galaxy.

\subsubsection{Physical properties and chemical abundances from the integrated spectrum of IIZw70}

In this section some of the more interesting spectroscopic properties
of the ionized gas in IIZw70, derived by analysing the integrated
spectrum are described. We obtained the integrated spectrum of IIZw70
by summing the emission from each fiber within an area of $\sim$ 300
arcsec$^{2}$ ($\sim$ 1.8 kpc$^{2}$), enclosing the whole nebular
emission.Integrated spectra were obtained for low and high spectral
resolution.

The analysis of the integrated spectrum presents several
advantages.  Contrary to previous studies that attempt to describe
the average ionization conditions in the ionized nebula using
individual spectra taken at different locations, we were able to
describe the average spectroscopic properties over an extended region. Another advantage is
that the integrated spectrum does not suffer from systematic effects 
that may be found for the spectra obtained with smaller apertures (Kewley et al. 2005). These
apertures are usually centered on bright SF knots, presenting a
bias towards young populations (as a matter of fact in most spectroscopic surveys of HII 
galaxies carried out up to now). Furthermore, one can
compare the real integrated properties of the local SF dwarf
galaxies with the properties of SF galaxies in
intermediate/high redshift for which only their integrated
characteristics are known due to their distance.

The emission lines corresponding to the integrated spectra were
measured following the same procedure as in Kehrig et al. (2004).
Reddening corrected line intensities, normalized to H$\beta$ are shown
in table~\ref{flux}, as well as the value of C(H$\beta$), estimated
using the H$\alpha$/H$\beta$ and H$\gamma$/H$\beta$ ratios, the
reddening corrected flux of H$\alpha$, F(H$\alpha$), EW(H$\beta$), EW(H$\alpha$),
and corrected velocity dispersion (see next section for details)
derived from H$\beta$ and [OIII]$\lambda$4959. The effect of the
underlying absorption in the hydrogen lines appears not to be
important; a 0.3 \AA~ EW correction for the underlying absorption on
the integrated spectrum would be enough to bring the observed Balmer
ratios to their theoretical value within the errors.

From the reddening corrected line intensities of the integrated
spectrum we derived ionization structure indicators, physical
conditions and chemical abundances (see table~\ref{tab4}). These
quantities were derived as described in section 3.4 for individual
fibers.

\begin{table}
\caption{Reddening corrected line intensities normalized to H$\beta$ for the
integrated spectrum of IIZw70. The reddening coefficient, C(H$\beta$), the reddening corrected flux of H$\alpha$ and H$\beta$, F(H$\alpha$) and F(H$\beta$)  in units 10$^{-13}$ erg cm$^{-2}$ s$^{-1}$, equivalent width of H$\alpha$ and H$\beta$, (EW(H$\alpha$) and EW(H$\beta$) in \AA), and the corrected velocity dispersions derived from H$\beta$ and [OIII]$\lambda$4959,
($\sigma_{cor}$(H$\beta$) and $\sigma_{cor}$([OIII]$\lambda$4959) in km s$^{-1}$), are included. Errors are quoted.}
\renewcommand{\footnoterule}{}
\label{flux}
\begin{minipage}{\textwidth}
\begin{tabular}{lccc} \hline
~~~$\lambda$~~~~ Ion   & flux  \\ \hline

3726 [OII] 	             &1.03   $\pm$0.01 \\
3729 [OII] 	             &1.38  $\pm$0.02 \\
3868 [NeIII]		    &0.31 $\pm$0.01 \\ 
3889 H8+HeI  	             &0.19 $\pm$0.01 \\  
3970 H$\epsilon$+[NeIII]     &0.08  $\pm$0.01 \\ 
4101 H$\delta$ 		     &0.24 $\pm$0.01  \\ 
4340 H$\gamma$ 		    &0.45 $\pm$0.01  \\  
4363[OIII] 		    &0.07 $\pm$0.01  \\  
4471 HeI		     &0.06 $\pm$0.01  \\ 
4861 H$\beta$ 		     &1.00 $\pm$0.02 \\
4959[OIII]                &1.20 $\pm$0.01  \\  
5007[OIII] 		  &3.76 $\pm$0.03  \\
6563 H$\alpha$ 	 &2.80 $\pm$0.03 \\
6584 [NII]                &0.17  $\pm$0.01 \\
6678 HeI 		  &0.03 $\pm$0.01 \\
6717 [SII] 		  &0.29  $\pm$0.01 \\
6731 [SII] 		  &0.20  $\pm$0.01 \\

\hline

C(H$\beta$)          &0.30      &            \\
F(H$\alpha$)         &10.72 $\pm$ 0.14    &       \\
F(H$\beta$) &3.83 $\pm$ 0.05\footnote{Assuming H$\alpha$/H$\beta$ = 2.80 for Case B recombination \\ (Storey \& Hummer 1995)}    &       \\
EW(H$\alpha$)(\AA)         &-221 $\pm$ 18    &           \\
EW(H$\beta$)(\AA)         &-44 $\pm$ 1    &           \\
$\sigma_{cor}$(H$\beta$) (km/s)    &19 $\pm$ 3     &          \\
$\sigma_{cor}$([OIII]$\lambda$4959) (km/s)    &21 $\pm$ 3     &          \\

\hline
\end{tabular}
\end{minipage}
\end{table}

\begin{table}
\caption{Physical conditions, abundances and ionization structure indicators  \label{tab4}}
\renewcommand{\footnoterule}{}
\begin{minipage}{\textwidth}
\begin{tabular}{lc} \hline
Parameter &Integrated \\ \hline
$T_e$(O {\sc iii})(10$^{4}$ K)         &1.46 $\pm$0.08 \\
$T_e$(O {\sc ii})\footnote{$T_{e}$([OII]) using $t_{e}$([OII])= 0.72$t_{e}$([OIII])+0.26 (Pilyugin et al. 2006)}(10$^{4}$ K) &1.31 $\pm$0.06 \\
$N_e$(O {\sc ii})(cm$^{-3}$)            &80:\footnote{Uncertain value} \\
$N_e$(S {\sc ii})(cm$^{-3}$)           &144 $\pm$27 \\
12+log O$^+$/H$^+$                     &7.40 $\pm$0.05 \\
12+log O$^{2+}$/H$^+$                  &7.62$\pm$0.05 \\
12+log (O/H)\footnote{O/H abundance derived using T$_{e}$[OIII]} &7.83 $\pm$0.04 \\
12+log(O/H)\footnote{O/H derived from P-method (Pilyugin \& Thuan 2005)} &7.99 $\pm$ 0.20 \\
12+log N$^+$/H$^+$                     &6.23$\pm$0.03 \\
log N/O                                &-1.17 $\pm$0.06 \\
12+log Ne$^{2+}$/H$^+$                 &6.96$\pm$0.06 \\
log Ne/O                               &-0.67 $\pm$0.07 \\
12+log S$^{+}$/H$^+$                   &5.79 $\pm$0.03 \\
10$^3$ He$^{+}$/H$^+$ ($\lambda$4471)            &76 $\pm$15 \\
10$^3$ He$^{+}$/H$^+$ ($\lambda$6678)            &85 $\pm$17 \\
log R$_{23}$\footnote{R$_{23}$ = ([OII]$\lambda$3726.29+[OIII]$\lambda$4959,5007)/H$\beta$}  &0.88 $\pm$0.03  \\
log [OIII]/[OII]\footnote{[OIII]$\lambda$4959,5007/[OII]$\lambda$3726.29}     &0.33 $\pm$0.01   \\
log [NII]/[OII]\footnote{(1.3 x [NII]$\lambda$6584)/[OII]$\lambda$3726.29}      &-1.04 $\pm$0.02         \\
log [SII]/H$\alpha$\footnote{[SII]$\lambda$6717.31/H$\alpha$} &-0.75 $\pm$0.01         \\

\hline
\end{tabular}
\end{minipage}
\end{table}


\subsection{Kinematics of the ionized gas}

One of the advantages to using two-dimensional spectroscopy is that
the spatial distribution of the kinematics of the ionized gas can be
studied when using high enough spectral resolution. Radial velocity
and line widths ($\sigma$ = FWHM/2.35) were obtained by fitting a
single Gaussian to the emission line. The width is corrected for the
instrumental profiles and for the thermal motions, following the
formula: 

\[\sigma_{cor}^{2} =  \sigma_{obs}^{2} - \sigma_{ins}^{2} - \sigma_{ther}^{2} \]

where $\sigma$$_{cor}$, $\sigma$$_{obs}$, $\sigma$$_{ins}$ and $\sigma$$_{ther}$ are
the corrected width, the observed width, the width of the instrumental
profile and the width due to the thermal motions, respectively.  The
$\sigma$$_{ins}$ is obtained from the profiles of emission lines of lamps, which
is $\sim$ 45 km s$^{-1}$ in our high dispersion spectra. To derive the width of the thermal
profile the electron temperature ($T_e$=14600K) derived from
the integrated spectrum is used as a representative value, finding a
value of $\sim$ 3 km s$^{-1}$ to be applied to the
[OIII]$\lambda$4959 emission line.

\begin{figure}[!ht]
 \mbox{
  \centerline{
\hspace*{0.0cm}\subfigure{\label{}\includegraphics[width=7cm]{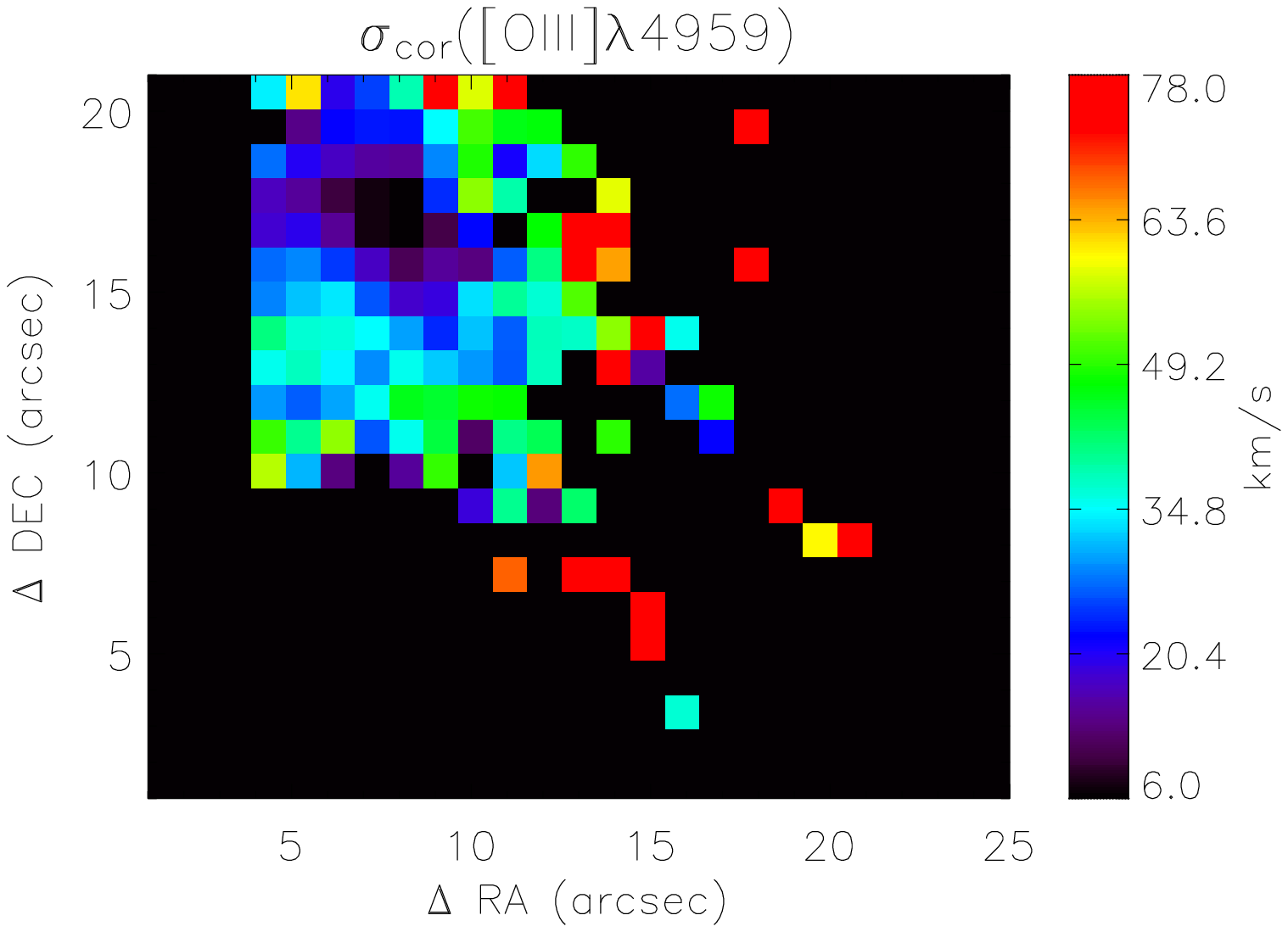}}
   }}
 \mbox{
  \centerline{
\hspace*{0.0cm}\subfigure{\label{}\includegraphics[width=7cm]{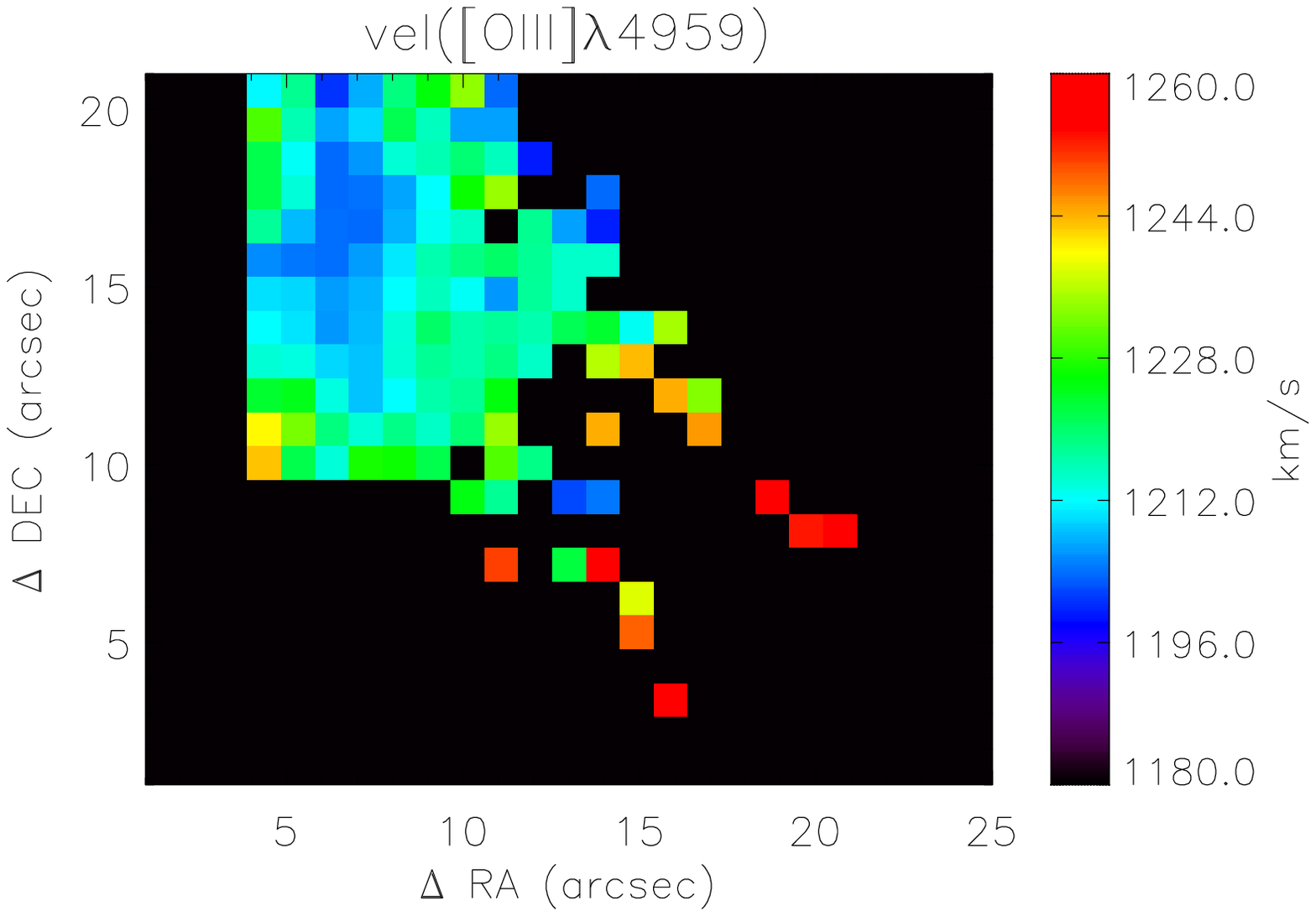}}
   }}
  \caption{Kinematics of IIZw70:  maps of the corrected velocity dispersion, $\sigma$([OIII]$\lambda$4959), and of the radial velocity as measured from [OIII]$\lambda$4959. The lowest values for both, radial velocity and velocity dispersion coincide with the central burst locus. North is up, east is left. Relative right ascension (RA) and declination (DEC) are shown in arcsec.}
\label{vel}
\end{figure}

In figure~\ref{vel} the maps of radial velocity and of corrected
velocity dispersion, $\sigma$$_{cor}$,  resulting from single
Gaussian fits to the [OIII]$\lambda$4959 emission line are shown. 

The corrected velocity dispersion $\sigma$$_{cor}$ map shows the
maximum values associated to the tails region of the galaxy, with
$\sigma$$_{cor}$ decreasing towards the brightest zones. This map also
shows a conspicuous minimum close to the position of the minimum
radial velocity. In addition, the corrected velocity dispersion values
measured from the integrated spectrum (see table~\ref{flux}) are
similar to those obtained for the brightest part of the galaxy (see
figure~\ref{vel}). This is because the velocity dispersion for HII
galaxies is expected to be dominated by the central (core) component
as shown by Telles et al. (2001). The map of the velocity field
observed presents a radial velocity minimum coincident with the locus
of the central burst.  The difference between this minimum and the
highest values of radial velocities, found in the tails to the
southwest, is of $\sim$ 60 km/s.\footnote{Cox et al. (2001) found a
value of 67 km/s for the rotational speed of the HI neutral gas.} This
blueshifted emission could be produced by an outflow from the bright
burst of the galaxy.  We assume the outflow interpretation because of
the double-tail morphology of the ionized gas and the $\sigma$$_{cor}$
increasing towards the tails region of the galaxy; whereas for a disk,
gas velocity dispersion is expected to fall with radius (Binney \&
Tremaine 1987, Bertola et al. 1995).

\section{Discussion and conclusions}

\subsection{Ionized gas and massive stars}

We analyzed integral field spectroscopic observations of the HII
galaxy IIZw70 in order to investigate the interplay between the
massive stars and the ionized gas. This galaxy is interacting with its
nearby companion IIZw71. The contiguous streamer of gas between IIZw70
and IIZw71, as well as the SF activity seen in both, indicate an
ongoing interaction and possible gas interchange between these two
galaxies (Cox et al. 2001). We studied the spatial distribution of
properties of the ionized gas (electron temperature and density,
chemical abundances, dust extinction, excitation, kinematics) from
different emission-line maps.  These maps reveal a central starburst
(Cair\'os et al. 2001, 2001) surrounded by an extended lower
excitation, low surface brightness ionized component. From our
integrated flux in H$\alpha$, corrected by extinction, we derived a
star formation rate of SFR(H$\alpha$) = 0.3 M$_{\odot}$/yr for IIZw70
following Kennicutt (1998); this value is in agreement with the value
we derived using the flux reported in Gil de Paz et al. (2003) from
CCD H$\alpha$ imaging. Kewley et al. (2002), using long-slit
spectroscopic mapping, derived a SFR(H$\alpha$) = 0.19
M$_{\odot}$/yr\footnote{We calculated this value converting the
SFR(H$\alpha$) given in Kewley et al. (2002) from the distance they
adopted to the distance adopted in this paper (see table
1).}. Different integration areas and C(H$\beta$) values could be
possible reasons for the discrepancy between our SFR(H$\alpha$)
measurement and the value found by Kewley et al. (2002).

For the first time we detected the presence of WR stars in IIZw70. For
the blue bump, our measurements of the equivalent width, EW(WRbump) =
-2.2 $\pm$ 0.6 \AA, and of the ratio 100 x I(WRbump)/I(H$\beta$) = 4.6
$\pm$ 1.1 (section 3.3) are in good agreement with the predictions of
the evolutionary synthesis models Starburst99 (assuming a Salpeter
IMF, M$_{upper}$ = 120 M$_{\odot}$ and Z=0.004 with high mass loss;
Leitherer et al. 1999). The models predict EW(WRbump) values around -2
\AA~ at $\sim$ 3.2 Myr after an instantaneous burst. Corresponding
predictions for 100 x I(WRbump)/I(H$\beta$) give values around 3.6
Myr. This fact indicates that the central burst in IIZw70 is very
young.  Assuming an age for the burst of 3.2 Myr (as deduced from our
measurements), we obtained a ratio of WR/O = 0.056 from
Starburst99. The presence of a bump at $\sim$ 5800 \AA~ could be an
indicator of the existence of WC stars. A broad feature at this
wavelength has been marginally detected in the same spectra, and
though the measured EW is consistent with the same model predictions,
its statistical significance is poor due to the low S/N. Overall, a
total number of $\sim$ 17 WN+WC stars are predicted by the models from
our findings.

The nebular emission of HeII$\lambda$4686 detected in this
work is shown to be spatially extended, though it is not coincident
with the location of the WR bumps by a few arcsec\footnote{We should
bear in mind, however, that a small contribution of nebular
HeII$\lambda$4686 could have remained unnoticed in the broad
feature}. This nebular HeII emission is seen to the southwest of the
box encompassing the probable ionizing clusters (see figure
4), suggesting the existence of very hard ionizing radiation which, in
principle, could not be necessarily related to the WR stars
located towards the northeast corner.

In the overall picture, IIZw70 is dominated by the presence of a young
central starburst with a significant SFR. The peak of the H$\alpha$
emission is located close to the probable youngest stellar clusters
(see figure~\ref{cumulo.WR.HeII}).  These young starburst could be
producing strong winds which have been able to develop an ionized gas
outflow from which we see the blueshifted component as a feature in
the radial velocity map of the galaxy (see figure 10). This outflow is
thought to have been responsible for the minimum in optical extinction
(slightly displaced from H$\alpha$ emission peak by
$\sim$2$^{\prime\prime }$ to the west) apparent in the
H$\alpha$/H$\beta$ map, where dust could have been swept up by the
wind. The characteristic size ($\sim$ 150 pc) and velocity ($\sim$ 60
km/s) of the outflow zone are consistent with previous findings in the
literature (see e.g. Roy et al. 1991) for the giant HII complex
NGC~2363). We can also see that the area with lower dispersion
velocity partially overlaps with the zone of the gas outflow, between
the probable ionizing clusters location and the zone of minimum
extinction. From the kinematical point of view, this picture appears
to be consistent with the velocity dispersion map. The narrowest line
profiles are dominated by the bulk of the emission from the central
HII regions, as expected (Telles et al. 2001), whereas the rest of the
ionized gas, with a lower surface brightness, seems to be suffering
the effects of the winds and its kinematics should be associated to
the structures generated by the likely interaction of the winds with
the ambient medium. This situation is reminiscent of what has been
found for other extragalactic HII regions and HII galaxies
(e.g. Mu\~noz-Tu\~n\'on et al. 1996, Telles et al. 2001) which host
strong starbursts.

\subsection{Chemical enrichment and abundance variations}

In this section, based on our chemical abundance measurements, we
present an analysis on the chemical enrichment of the ionized gas and
chemical abundance variations in IIZw70.

The oxygen abundance derived here for the integrated spectrum of
IIZw70 is 12 + log(O/H) = 7.83 $\pm$ 0.04. Previous works reported
direct values of oxygen abundance for IIZw70 using long-slit spectra
positioned in different places across the brightest regions of
IIZw70. Lequeux et al. (1979) quoted a value of 8.07; Kobulnicky \&
Skillmann (1996) derived an oxygen abundance of 8.07 $\pm$ 0.08
whereas Shi et al. (2005) found 12+log (O/H) = 7.69. In principle, the
differences between our O/H measurement and the values found in the
literature could be explained by different integration areas (see
figure 9, bottom panel; Kewley et al. 2005) and atomic parameters
used.

Thus far, in most works on chemical abundances in dwarf galaxies there
is no clear evidence of abundance variations. Though selected examples
of nitrogen enhancement in dwarf galaxies have been reported in the
literature, notably the case of NGC~5253 (e.g. L\'opez-S\'anchez et
al. 2006 and references therein), this is not the case of O/H. Many
works in the literature have searched for a possible oxygen abundance
variation within dwarf galaxies. Kobulnicky \& Skillman (1996)
presented optical spectroscopic chemical measurements of the ISM in
the irregular galaxy NGC 1569 that reveal no substantial localized
chemical self-enrichment (i.e., ``local pollution''; see e.g. Kunth \&
Sargent 1986; Pagel, Terlevich \& Melnick 1986; Pilyugin
1992). V\'{\i}lchez \& Iglesias-P\'aramo (1998), using bi-dimensional, long-slit
spectroscopy, showed that IZw18 presents a substantially homogeneous
chemical composition over the whole galaxy. 
From long-slit spectra of 16 HII regions in the dwarf irregular NGC 1705, Lee \& Skillman
(2004) reported that there is no significant spatial variation of
oxygen abundances for this galaxy.
Izotov et al. (2006), from two-dimensional spectroscopy of the HII galaxy
SBS0335-052E, found a variation of 0.4 dex in the oxygen
abundance. They concluded that these variations may not be
statistically significant due to unaccounted error estimates.  Lee et
al. (2006) obtained a slightly significant (3.2$\sigma$) oxygen
abundance gradient of -0.16 $\pm$ 0.05 dex kpc$^{-1}$ within the Local
Group dwarf irregular galaxy NGC~6822. However they claim further deep
high-quality spectra of nebulae and stars are needed to distinguish
clearly between either a zero or a nonzero slope.

In this work we calculated the O/H abundance in 16 positions across
IIZw70 directly from the measurement of [OIII]$\lambda$4363. From
these values we found a maximum variation in the derived oxygen
abundances of $\sim$ 0.4 dex, similar to the range reported by Izotov
et al. (2006) for SBS0335-052E. In order to study whether this
variance is statistically significant for an effective abundance
variation we must include all other unaccounted uncertainties
(variable seeing, flux calibration, etc), which surely contribute to
the total oxygen abundance error but are difficult to estimate. In any
case, our observations indicate that, within $\sim$ $\pm$ 0.2 dex, the
ionized gas in the central burst of IIZw70 appears to be chemically
homogeneous in O/H over spatial scales of hundreds of parsecs. In the
case of the neon-to-oxygen ratio, Ne/O, the range of values as derived
from our point by point abundance analysis, does not show
inhomogeneity within the errors; as is also the case of the
nitrogen-to-oxygen ratio, N/O (see section 3.4).  All these findings
give us the upper limits to any chemical inhomogeneity in the ionized
gas in the brightest part of IIZw70. In accordance with these results
we can say that we are probably not detecting a ``local pollution''
case in IIZw70, consistent with most results obtained by other works
cited previously.

A favourable explanation to this apparent degree of homogeneity in
dwarf galaxies is given by the scenario presented in Tenorio-Tagle
(1996). According to this scenario, in those places in which one
expects that newly produced metals could be returned to the ISM, the
``local pollution'' is not detected in the warm phase of the
ISM. Rather the newly synthesized metals are returned into the hot
phase of the ISM, and there is a significant time delay before these
newly produced metals can be detected in the warm phase of the
ISM. Variations of the oxygen abundance in the ISM of HII galaxies
could give constraints to the physical mechanisms involved in the
recycling processes (e.g. Recchi et al. 2001).

Analysing the predictions of chemical evolution models, we verified
that continuous SF models (Moll\'a \& D\'\i az 2005) predict an
enhancement in O/H of up to 0.4 dex, to be produced in spatial scales
of $\sim$ 1 kpc along their disk, over timescales of 8 Gyr. The
important point here is that these models predict O/H enhancements
that could be achieved keeping the N/O ratio approximately constant
throughout the process, and at an absolute value consistent with our
observations (see section 3.4) to within 0.1 dex. Further observations
with IFUs in 10 m class telescopes and additional measurements of
chemical abundances in IIZw70, covering a larger area, should be
required in order to draw a firmer conclusion on the statistical
contribution of local variations to the chemical homogeneity.  Given
the interest in the problem of chemical homogeneity of dwarfs galaxies
and its role in their chemical evolution (e.g. Legrand et al. 2001,
Recchi et al. 2001) a clearer view of the observational situation is
probably warranted.

\begin{acknowledgements}

C.K wishes to acknowledge Consejo Superior de Investigaciones
Cient\'{\i}ficas (CSIC-Spain) for an I3P fellowship; C.K. also thanks
the CAPES (Brazil) for a fellowship and the Observatorio Nacional for
continuous support.  This research was partially funded by projects
AYA2004-08260-C03-02 ``Estallidos'' of the Spanish PNAYA and TIC~111
of the Junta de Andalucia (Spain). S.F.S as part of the research group
FQM322 would like to thank the Spanish Plan Nacional de Astronom\'\i a
program AYA2005-09413-C02-02, of the Spanish MEC and the Plan Andaluz
de Investigaci\'on of Junta de Andaluc\'{\i}a. We wish to thank the
anonymous referee for his/her useful comments and suggestions. We
thank M. Villar-Mart\.in and D. Reverte-Paya for their help in the
initial stages of this project. During the course of this work, we
benefited from conversations with M. Moll\'a and V. Luridiana. Thanks
are due to G. Stasi\'nska for her fruitful comments and careful
reading of the manuscript. We also want to thank Steve Donegan for his
careful revision of the English in this paper.

\end{acknowledgements}

\appendix

\end{document}